\documentclass[singlecolumn,superscriptaddress]{revtex4}
\usepackage[utf8]{inputenc}
\usepackage{empheq}
\usepackage{amsmath}
\usepackage{amsfonts}
\usepackage{amssymb,bm,txfonts}
\usepackage{mathtools}
\usepackage{stmaryrd}
\usepackage{dsfont}
\usepackage{graphicx}
\usepackage[bookmarksnumbered=true]{hyperref}
\usepackage{stackrel}
\usepackage{color}
\usepackage{comment}
\usepackage{xcolor}
\usepackage{enumitem}
\usepackage{moresize}
\usepackage{graphicx}
\usepackage{siunitx}
\usepackage{fancyhdr}

\def \ee{\end{equation}}
\def \be{\begin{equation}}
\def \eea{\end{eqnarray}}
\def \bea{\begin{eqnarray}}

\begin{document}
\title{Symmetries of Quantum Spacetime:\\ 
%Lie Derivatives and 
Conserved Quantities
within the $q$-desic Framework}
\author{Benjamin Koch}
\email{benjamin.koch@tuwien.ac.at}
%\affiliation{Institut f\"ur Theoretische Physik und Atominstitut,
% Technische Universit\"at Wien,
% Wiedner Hauptstrasse 8--10,
% A-1040 Vienna, Austria}
\affiliation{Institut f\"ur Theoretische Physik, Technische Universit\"at Wien, Wiedner Hauptstrasse 8-10, A-1040 Vienna, Austria}
\affiliation{Atominstitut, Technische Universit\"at Wien,  Stadionallee 2, A-1020 Vienna, Austria}
 \affiliation{Pontificia Universidad Cat\'olica de Chile \\ Instituto de F\'isica, Pontificia Universidad Cat\'olica de Chile, \\
Casilla 306, Santiago, Chile}
\author{Ali Riahinia}
\email{ali.riahinia@tuwien.ac.at}
\affiliation{Institut f\"ur Theoretische Physik, Technische Universit\"at Wien, Wiedner Hauptstrasse 8-10, A-1040 Vienna, Austria}
\affiliation{Atominstitut, Technische Universit\"at Wien,  Stadionallee 2, A-1020 Vienna, Austria}

\author{Ángel Rincón}
\email{angel.rincon@physics.slu.cz}
\affiliation{Departamento de F{\'i}sica, Universidad del B{\'i}o-B{\'i}o, Casilla 5-C, Concepci{\'o}n, Chile}
\affiliation{
Research Centre for Theoretical Physics and Astrophysics, Institute of Physics, Silesian University in Opava,
Bezručovo náměstí 13, CZ-74601 Opava, Czech Republic.
%Departamento de Física Aplicada, Universidad de Alicante, Campus de San Vicente del Raspeig, E-03690 Alicante, Spain
}

\begin{abstract}
Classical spacetime symmetries can be characterised in terms of Killing vectors. When combined with geodesic motion,
these vectors allow to 
identify the conserved quantities along particle trajectories. 
In quantum spacetime, however, an effective motion of test particles can be encoded by $q$-desics. This generalisation of the geodesic equation is governed by expectation values of quantised geometric operators.
In this framework, the classical correspondence between symmetries $\leftrightarrow$ Killing vectors $\leftrightarrow$ conserved quantities needs to be revisited.

In this paper we establish this relation within the quantum gravitational $q$-desic framework. We generalise the Killing vectors to this framework and derive the associated conserved quantities. 
%This provides a systematic characterisation of symmetries directly at the level of effective quantum-gravitational trajectories. 
We further analyse the reparametrisation gauge freedom of the $q$-desic equations and clarify the role of affine and non-affine parametrisations. Particular attention is devoted to the evolution of the four velocity norm. 
%While metric geodesics preserve $u^2$, $q$-desic corrections can generically drive $u^2$ away from its initial value, providing a geometric criterion for effective mass non-conservation and, for initially null trajectories, for departures from the classical light cone. 
Our results establish a symmetry framework for $q$-desics and extend the classical link between spacetime invariance, conserved quantities, and particle motion into the quantum-gravitational regime.
\end{abstract}
\maketitle
\tableofcontents
%%%%%%%%%%%%%%%%%%%%%%%
\section{Introduction} \label{intro}

%%%%%%%%%%%%%%%%%%%%%%
\subsection{From Quantum Gravity to particle motion}

General Relativity (GR)~\cite{Einstein:1916vd,Einstein:1915ca} and Quantum Theory (QT) are the two foundational pillars of modern theoretical physics. On the one hand, GR provides a geometric description of gravitation in terms of a dynamical spacetime metric. An early demonstration of the structure and predictive power of Einstein's field equations was provided by Schwarzschild, who obtained the first exact vacuum solution describing the exterior gravitational field of a spherically symmetric mass~\cite{Schwarzschild:1916uq}. This solution later became fundamental to the relativistic description of compact objects and black-hole spacetimes. GR has been confirmed over a wide range of scales, from solar-system tests to strong-field gravitational phenomena and cosmology~\cite{Will:2014}. 
More recently, this
confirmation has been extended to the genuinely strong-field, dynamical
regime through the direct detection of gravitational waves from compact
binary coalescences~\cite{LIGOScientific:2016aoc} and through
very-long-baseline interferometric imaging of the shadow of the
supermassive black holes at the centers of M87 and the Milky
Way~\cite{EventHorizonTelescope:2019dse,EventHorizonTelescope:2022wkp},
which together provide some of the most stringent tests of GR to date.
On the other hand, QT underlies the successful description of matter and interactions at microscopic scales~\cite{Heisenberg:1925,Schrodinger:1926,vonNeumann:1927} and forms the conceptual basis of quantum field theory~\cite{Dirac:1927}. Despite their outstanding empirical success, the conceptual and mathematical frameworks of GR and QT are profoundly different. GR is a classical theory with a dynamical causal structure, whereas QT is formulated in terms of states, operators, superposition, and Hilbert-space structures~\cite{vonNeumann:1927}. Einstein was among the first people to argue that quantum effects should modify his theory of general relativity~\cite{Einstein:632328}. 
Since then, a large number of alternative research programs have been used to achieve this reconciliation from different starting points, 
such as for example: (i) string/M-theory~\cite{Polchinski:1998rq}, (ii)
canonical QG~\cite{Thiemann:1992jj}, (iii)
effective field theory of gravity~\cite{Donoghue:1994dn,Calmet:2017qqa}, (iv)
loop quantum gravity and its associated spin-foam formulations~\cite{Rovelli:2004tv,Ashtekar:2004eh}, (v) causal dynamical triangulations~\cite{Ambjorn:2012jv}, (vi) asymptotically safe gravity~\cite{Reuter:2012id}, (vii) scale-dependent gravity \cite{Koch:2016uso,Rincon:2017goj} and (viii) approaches based on non-commutative geometry, in which the classical smooth manifold structure of spacetime is replaced by an algebra of non-commuting operator-valued coordinates~\cite{Szabo:2001kg,Connes:1994yd}. Albeit these programs aim to reproduce GR in some limit, their underlying principles and predictions may differ strongly at a deeper level. 
Which of these approaches, if any, correctly describes nature is unknown. This question can only be answered by a comparison to observation and experimental measurements.
The $q$-desic framework offers a new tool for such a comparison, which will be complementary to e.g. the envisaged observation of single gravitons~\cite{Shenderov:2024rup}.

Symmetry principles play a central role in both theories. In GR, spacetime symmetries are encoded geometrically through diffeomorphisms, Lie derivatives, and Killing vector fields~\cite{Einstein:1916vd,Slebodzinski:1931,Killing:1892}. They organise the structure of exact solutions and lead to conserved quantities such as energy and angular momentum~\cite{Noether:1918}. In QT, classical symmetries can either be maintained~\cite{Noether:1918,Stone:1932,Wigner:1939} or broken. In the latter case one speaks of quantum ``anomalies''.
A priori, both types of scenarios could occur in the context of $q$-desic motion.

As it turns out, it is possible to construct a $q$-deformed generalisation of the Lie derivative acting on operator-valued tensor fields within the $q$-desic framework. Based on this, we derive the corresponding deformed Killing equation, and we show how it gives rise to $q$-deformed conserved quantities along $q$-desics that reduce smoothly to their classical counterparts, energy and angular momentum among them, in the proper limit. This paper is organised as follows.
In Sec.~\ref{intro} we briefly outlined the context of the problem to be
addressed, as well as the research question of the paper.
Then, in Sec.~\ref{Setting} we summarize the technical elements of
the paper, including the definition of the Killing vector and both the
classical geodesic equation and its corresponding $q$-desic counterpart.
Finally, radial canonical quantum gravity (with static equal-radius
quantization) is properly described.
Subsequently, in Sec.~\ref{form} the formalism is described in
detail, focusing on the expectation values of the quantities involved.
Conserved quantities are investigated in Sec.~\ref{conserved}, and a
concrete case study in quantum spacetime is summarized in
Sec.~\ref{caseofstudy}.
Before concluding, in Sec.~\ref{disc} we discuss our main findings, which
include how proper time affects our results within this formalism, gauge
invariance, and details of the $q$-Killing equation, among other topics.
Finally, we present our conclusions in the last section.

%Since these questions are most transparent when contrasted with their classical counterparts, we begin by reviewing the role of symmetries, Killing vectors, geodesics, and conserved quantities in the classical theory.

%%%%%%%%%%%%%%%%%%%%%%
\subsection{Research question
%of the paper
}

%One of the pillars of GR is the consistent link between the {\it Killing symmetries} of a given metric tensor and the {\it conserved quantities along a geodesic curve}~\cite{Castellani:1981us,Gitman:2004qs}.
%We explore the question {\it whether or how this relation still holds in the context of quantum gravity}.
%In particular, we will focus on the expectation values of metric operators as they appear in the $q$-desic equation. 
%\textcolor{red}{??? Ali: These two papers are called "Symmetries in constraint Hamiltonian systems"\cite{Castellani:1981us} and "Symmetries in Constrained Systems"\cite{Gitman:2004qs} !! maybe it would be nice to mention it here. Its a 1982 paper and it could be a good reference. ??? }
One of the pillars of GR is the close relation between {\it Killing symmetries} of the spacetime metric and {\it conserved quantities along geodesic motion}. Related aspects of symmetries and conserved charges in constrained Hamiltonian systems have been discussed in Refs.~\cite{Castellani:1981us,Gitman:2004qs}. Here, we investigate whether and how the Killing-symmetry/conservation correspondence persists in the context of quantum gravity, focusing in particular on expectation values of metric operators entering the $q$-desic equation.

%%%%%%%%%%%%%%%%%%%%%%%%%%%%%
\section{Setting the stage} \label{Setting}

In this section we will summarise the physical and mathematical toolkit that will be used in this paper.

\subsection{Spacetime Symmetries in General Relativity}

%The Lie derivative of a metric tensor in general relativity allows us to determine the symmetries and conserved quantities of the system by solving the killing equation.\\
%For the metric tensor $g_{\mu \nu} (x)$, the Lie derivative reads
%\begin{equation}
%    \left( \mathcal{L}_{\xi} g \right)_{\mu \nu} = \xi^{\rho}  \partial_{\rho} g_{\mu \nu} +  g_{\mu \beta} \partial_{\nu} \xi^{\beta} +  g_{\alpha \nu}  \partial_{\mu} \xi^{\alpha}  = \nabla_{\mu} \xi_{\nu} + \nabla_{\nu} \xi_{\mu} = \partial_{\mu} \xi_{\nu} + \partial_{\nu} \xi_{\mu} - 2 \Gamma^{\alpha}_{\mu \nu} \xi_{\alpha} 
%\end{equation}
%Killing vectors are those with a vanishing Lie derivative
%\begin{equation}
%    \left( \mathcal{L}_{\xi} g \right)_{\mu \nu} = 0
%\end{equation}

Spacetime symmetries in General Relativity are described by transformations that preserve the geometric structure encoded in the metric tensor. Infinitesimally, such transformations are generated by vector fields, and their action on the metric is measured by the Lie derivative~\cite{Slebodzinski:1931}. This provides the natural starting point for defining Killing vector fields and relating spacetime symmetries to conserved quantities~\cite{Killing:1892,Noether:1918}. In General Relativity, the geometry of spacetime is described by a smooth manifold $\mathcal{M}$ equipped with a metric tensor $g_{\mu\nu}$~\cite{RicciLeviCivita:1900,Einstein:1916vd}.

Consider the infinitesimal transformation
\begin{equation}
x'^{\mu}=x^{\mu}+\epsilon\,\xi^{\mu}.
\end{equation}
To first order in $\epsilon$, the transformed metric components are
\begin{equation}
g'_{\mu\nu}(x')
=
g_{\mu\nu}(x)
-\epsilon\left(
g_{\rho\nu}\partial_{\mu}\xi^{\rho}
+g_{\mu\rho}\partial_{\nu}\xi^{\rho}
\right),
\end{equation}
whereas
\begin{equation}
g_{\mu\nu}(x')
=
g_{\mu\nu}(x)
+\epsilon\,\xi^{\rho}\partial_{\rho}g_{\mu\nu}.
\end{equation}
The Lie derivative is therefore defined by~\cite{Slebodzinski:1931}
\be
(\mathcal{L}_{\xi}g)_{\mu\nu}
\equiv
\lim_{\epsilon\rightarrow 0}
\frac{g_{\mu\nu}(x')-g'_{\mu\nu}(x')}{\epsilon}
=
\xi^{\rho}\partial_{\rho}g_{\mu\nu}
+g_{\rho\nu}\partial_{\mu}\xi^{\rho}
+g_{\mu\rho}\partial_{\nu}\xi^{\rho}
=
\nabla_{\mu}\xi_{\nu}+\nabla_{\nu}\xi_{\mu}.
\ee

A vector field $\xi^{\mu}$ is called a Killing vector field if it leaves the metric invariant, resulting in a vanishing Lie derivative~\cite{Killing:1892},
\begin{equation}\label{eq_Killing}
    (\mathcal{L}_{\xi} g)_{\mu\nu} = 0 \; \longrightarrow \; \nabla_\mu \xi_\nu + \nabla_\nu \xi_\mu = 0 .
\end{equation}
This equation is also called the Killing equation. Its solutions characterise the continuous isometries of the spacetime~\cite{Killing:1892}. For the Schwarzschild spacetime, spherical symmetry gives rise to three rotational Killing vectors, while the static character of the spacetime provides an additional Killing vector associated with time translations. The resulting isometry group is $\mathbb{R}\times SO(3)$~\cite{Schwarzschild:1916uq,Jebsen:1921,Eiesland:1925}. A convenient basis for these Killing vector fields is
\begin{equation}
    \xi_{(0)} = \partial_t
\end{equation}
\begin{equation}
    \xi_{(1)} = \partial_{\phi}
\end{equation}
\begin{equation}
    \xi_{(2)} = - \cos (\phi) \partial_{\theta} +\sin (\phi) \cot (\theta) \partial_{\phi}
\end{equation}
\begin{equation}
    \xi_{(3)} = \sin (\phi) \partial_{\theta} + \cos (\phi) \cot (\theta) \partial_{\phi}
\end{equation}

The importance of Killing vectors is twofold. First, they provide a precise geometric characterisation of spacetime symmetry. Second, they are directly related to conserved quantities for particle motion~\cite{Noether:1918}. For geodesic motion, if a spacetime admits a timelike Killing vector, the corresponding contraction of the particle four-momentum with the Killing vector is conserved and can be identified with the conserved energy. Similarly, rotational Killing vectors give rise to conserved components of angular momentum. In this way, spacetime symmetries reduce the effective dynamical complexity of the problem and provide physically meaningful constants of motion~\cite{Noether:1918}.

More generally, the Lie derivative measures the change of a tensor field along the flow generated by $\xi^{\mu}$~\cite{Slebodzinski:1931}. The condition that the Lie derivative of the metric vanishes therefore expresses the statement that the geometry is invariant under the corresponding infinitesimal transformation. In classical GR, this provides the standard link between geometry, symmetry, and conservation laws~\cite{Killing:1892,Noether:1918}.

%%%%%%%%%%%%
\subsection{Geodesics and conserved quantities in General Relativity}

The motion of a free test particle in GR is described by a geodesic of the spacetime metric~\cite{Einstein:1916vd,Poisson:2011nh}. A convenient starting point is the point-particle action~\cite{Poisson:2011nh} e.g. in its Hamiltonian form
%\be\label{eq_SHGR}
% S_H(x^\mu, p^\mu, g_{\mu \nu})= \int d\lambda \left(p^\mu  g_{\mu \nu}\dot{ x}^\nu-\frac{\omega}{2} \left(p^\mu  g_{\mu \nu}p^\nu-m^2 \ \right)\right).
%\ee
\be \label{eq_SHGR}
S = \int d \lambda
\sqrt{\frac{dx^\mu}{d\lambda} \frac{dx^\nu}{d\lambda} g_{\mu \nu}}
\ee
where $\lambda$ is a  parameter along the worldline. Equivalently, the particle
dynamics can be formulated in a constrained Hamiltonian form, in which the
freedom to reparametrise the worldline is reflected by an arbitrary multiplier
(or gauge function)~\cite{Dirac:1950pj}.  Varying the action with respect to the dynamical canonical pair $x^\mu$ and $p^\mu$ and using the proportionality  $dx^\mu /d\lambda = \omega p^\mu$ yields the geodesic equation 
\begin{equation}\label{eq_Geodesic1}
    \frac{d^2 x^\mu}{d\lambda^2}
    = - \Gamma^\mu_{\alpha\beta}
    \frac{dx^\alpha}{d\lambda}
    \frac{dx^\beta}{d\lambda}
    - u^\mu \frac{\dot \omega }{\omega} ,
\end{equation}
where the $\omega$ is a free gauge function reflecting the aforementioned proportionality factor. 
Here, the connection $\Gamma^\mu_{\alpha\beta}$ is the Levi–Civita connection associated with the metric $g_{\mu \nu }$ whose coefficients are the Christoffel symbols $\Gamma^\mu_{\alpha\beta}=\left \{ ^\mu_{\alpha \beta}\right\}$. This connection follows directly from the variation of the point-particle action given in eq.~\ref{eq_SHGR}.
%Here, we assume metricity such that the connection is given in terms of the Levi-Civita symbol $\Gamma^\mu_{\alpha\beta}=\left \{ ^\mu_{\alpha \beta}\right\} $.
At this point it is useful to define some notation with respect to the parameters that we may use to describe a geodesic curve. We recall the definitions: 
\begin{itemize}
    \item $\lambda$: Parameter that is used to describe a curve.
    \item $\omega$: gauge function that represents the freedom of choice for the parameter $\lambda$. Here, we use $\dot \omega$ as short notation for $d\omega/d\lambda$. 
    If $\dot \omega =0$, $\lambda$ shall be called an affine parameter.
    \item $dt$: Time parameter of a given coordinate system. In most cases, the time parameter is not affine, meaning that $(d\omega)/dt \neq 0$.
    \item $d\tau$: Stands for the proper length in a given metric $ds^2=dx^\mu dx^\nu g_{\mu \nu}$. The proper length is not an affine parameter if  $(d\omega)/ds \neq 0$. 
\end{itemize}

Note, that 
for convenience, the parameter $\lambda$ is typically chosen to be affine such that the  gauge function $\omega$ is constant. For now we will adopt this condition, but later on we will return to the subtleties of this gauge choice.

The equation (\ref{eq_Geodesic1}) expresses the free-fall motion of a test particle in curved spacetime~\cite{Einstein:1916vd,Droste:1917}. The geodesic equation is directly connected to conserved quantities whenever the background admits symmetries~\cite{Killing:1892,Noether:1918}. Let $u^\mu = dx^\mu/d\lambda$ be the tangent vector to the geodesic and let $\xi^\mu$ be a Killing vector field. Then the quantity
\be\label{eq_conserved1}
C=   \sigma_i\omega \,\xi_\mu u^\mu.
\ee
is conserved along the motion where $\sigma_i$ is an integration constant. Indeed, using the geodesic equation and the Killing equation, one finds
\begin{equation}
    \frac{d}{d\lambda}C=\frac{d}{d\lambda}\left( \sigma_i \omega \xi_\mu u^\mu\right) = \frac{d x^{\nu}}{d \lambda } \frac{d}{d x^{\nu}}\left(\sigma_i \omega \xi_\mu u^\mu\right) 
    = \sigma_i u^\nu \nabla_\nu \left(\omega \xi_\mu u^\mu \right) = \sigma_i \left( \omega u^\mu u^\nu \nabla_\nu \xi_\mu + \omega \xi_\mu u^\nu \nabla_\nu u^\mu + u^\nu \xi_\mu u^\mu \nabla_{\nu} \omega \right) = 0
\end{equation}
where we have also used $ \partial_{\nu} C = \nabla_{\nu} C$ since $C$ is a scalar. Therefore,
every Killing symmetry gives rise to a constant of motion along a
geodesic~\cite{Killing:1892,Noether:1918}

For static and spherically symmetric spacetimes, these conserved quantities take their standard form. Time-translation symmetry implies a conserved energy, while rotational symmetry implies conserved angular momentum~\cite{Droste:1917,Hagihara:1931}. Restricting, without loss of generality, to the equatorial plane $\theta=\pi/2$, one typically obtains
\begin{equation}\label{eq_conservedCl}
    E = g_{tt}\,\dot t,
    \qquad
    l = r^2 \dot\phi ,
\end{equation}
where a dot denotes differentiation with respect to the affine parameter. Together with the normalisation condition
\begin{equation}
    g_{\mu\nu} u^\mu u^\nu =
    \begin{cases}
        1, & \text{timelike geodesics},\\
        0, & \text{null geodesics},
    \end{cases}
\end{equation}
these relations reduce the problem of motion to an effective radial equation~\cite{Droste:1917,Hagihara:1931}.

This classical interplay between geodesics, Killing symmetries, and conserved quantities is one of the central reasons why symmetries are so powerful in gravitational physics~\cite{Killing:1892,Noether:1918}. It also makes clear why these concepts must be revisited in a quantum gravity. Once the metric itself becomes operator-valued, the status of Lie derivatives, Killing equations, and conserved quantities can no longer be assumed to remain identical to their classical form.
%%%%%%%%%%%%%%
\subsection{$q$-desic: particle motion in QG} \label{sec_qdesics}

The quantum version of the geodesics equation with quantised metric operator was derived in Ref.~\cite{Koch:2025qzv}. The starting point was chosen to be the operator valued action
\be
\label{eq_Shat}
\hat S = \int d \lambda
\sqrt{\frac{dx^\mu}{d\lambda} \frac{dx^\nu}{d\lambda}\hat g_{\mu \nu}}
\ee
The idea of the $q$-desic approach is to perform all operations at the operator level
and only at the final step recur to the metric EV.
A variation with respect to $\delta x^\mu$ yields after bringing all operator valued expressions to the right hand side and then taking the QG Expectation Value (EV)
\be \label{eq_eomhatx}
\frac{d^2 x^\mu}{d\lambda^2}
=-
\langle \hat \Gamma^{\mu}_{\alpha \beta} \rangle
\frac{dx^\beta}{d\lambda} 
\frac{dx^\alpha}{d\lambda},
\ee
which is the version of the $q$-desic equation that was derived in Ref.~\cite{Koch:2025qzv}.
Note that some aspects of (\ref{eq_eomhatx}) were also proposed based on different arguments~\cite{Pipa:2018bui,Kanatchikov:2025tlm,Piazza:2022amf,Dalvit:1999wd,Parikh:2020kfh}.
Further, a complementary viewpoint to this is to consider the quantum properties of the velocities $dx^\mu/d\lambda$ and their entanglement~\cite{Balsells:2024lpl,Balsells:2025iac}.
We can add one interesting aspect to the relation (\ref{eq_eomhatx}) by noting that the action (\ref{eq_Shat}) is invariant under the local gauge transformations
\be
\lambda \rightarrow \tilde \lambda (\lambda).
\ee
After such a gauge transformation with a monotnous function,
the $q$-desics equation reads
\be\label{eq_qdesic}
\frac{d u^\mu}{d\lambda}
=-\left\langle 
\hat \Gamma^\mu_{\alpha \beta}
\right\rangle u^\alpha u^\beta - u^\mu \frac{\dot \omega}{\omega},
\ee
where we defined $\omega =d\tilde \lambda/d\lambda$ and
$u^\mu=dx^\mu /d\lambda$.
The difference between the $q$-desic (\ref{eq_qdesic})
and the geodesic (\ref{eq_Geodesic1}) arises from the fact that $\left\langle 
\hat \Gamma^\mu_{\alpha \beta}
\right\rangle \neq 
 \Gamma^\mu_{\alpha \beta}$. However we can quantify this difference by writing
 \be\label{eq_Covariance}
 \left\langle 
\hat \Gamma^\mu_{\alpha \beta}
\right\rangle= \Gamma^\mu_{\alpha \beta}(\langle \hat g \rangle)
+ \Delta^\mu_{\alpha \beta}.
 \ee
Only if all components of $\Delta^\mu_{\alpha \beta}$ vanish, the $q$-desic and the geodesic are identical. Thus, when we deal with $q$-desics and their solutions, it is mathematically analogous to working 
with ``conventional'' non-metricity~\cite{Hehl:1994ue}, where the connection contains a Levi-Civita part and some deviations 
\be
\Gamma^\mu_{\alpha \beta}
= \left \{ ^\mu_{\alpha \beta}\right\} + \Omega^\mu_{\alpha \beta}.
\ee
In this analogy, we can assign the following roles:
\bea\label{eq_Analogy}
\left\langle 
\hat \Gamma^\mu_{\alpha \beta}
\right\rangle &\leftrightarrow & \Gamma^\mu_{\alpha \beta},\\ \nonumber
\Gamma^\mu_{\alpha \beta}(\langle \hat g \rangle)
&\leftrightarrow & 
\left \{ ^\mu_{\alpha \beta}\right\},\\ \nonumber
\Delta^\mu_{\alpha \beta}
&\leftrightarrow & 
\Omega^\mu_{\alpha \beta}.
\eea
Thus, there may even exist an effective mapping to a description based on a variational principle in terms of Finsler geometry~\cite{Heisenberg:2026ess,Csillag:2026kdc} and statistical manifolds~\cite{Iosifidis:2023wbx}.
For the definition of conserved charges in parallel theories of gravity see~\cite{Koivisto:2026yxt,Koivisto:2026rmp}.\\
However, when we invoke such a formal analogy, we have to be aware that the $\Delta^\mu_{\alpha \beta}$ arise from taking the expectation values of products of metric operators, as it can be shown explicitly for the case of static space-times with spherical symmetry~\cite{Koch:2025yuz}.
For such spherically symmetric systems, we can write the $q$-desic equation
(\ref{eq_qdesic}) as
\be\label{eq_GeodSym}
\left(
\begin{array}{c}
   \dot  u^t  \\
   \dot  u^r\\
   \dot  u^\theta\\
   \dot  u^\phi
\end{array}
\right)
=
\left(
\begin{array}{c}
-\langle \hat g'/(\hat g)\rangle u^r u^t-\frac{\dot \omega}{\omega}u^t\\
(u^r)^2 \langle\hat g'/(2 \hat g)\rangle+\left\langle\hat g\left(r (u^\theta)^2+r \sin^2\theta (u^\phi)^2-  \hat n^2 \hat g' (u^t)^2/2 \right)\right\rangle-\frac{\dot \omega}{\omega}u^r\\
- 2 u^r u^\theta /r + \cos \theta \sin \theta (u^\phi)^2-\frac{\dot \omega}{\omega}u^\theta\\
- 2(u^r u^\phi+ r \cot \theta u^\theta u^\phi)/r-\frac{\dot \omega}{\omega}u^\phi
\end{array}
\right).
\ee
Here, we used a ``restricted Weyl
ordering prescription'', which means that one allows the cancellation of  operator functions with their respective inverse, before one imposes a Weyl ordering and evaluates expectation values.
Note that the operators $\langle \hat \Gamma^t_{rt}\rangle= \langle \hat \Gamma^r_{tt}\rangle$ are potentially divergent due to the appearance of $\langle 1/ \hat g\rangle$. This divergence can be treated in different ways as we discuss in the appendix~\ref{sec_divergencies}. Thus, whenever we use  expectation values of these operators in the subsequent discussions, we mean their finite part.
It is further worthwhile noting an interesting consequence of (\ref{eq_GeodSym}):
Once an equatorial plane for a test particle is chosen by some initial condition, e.g. $u^\theta =0$ (or $u^\phi=0$), all subsequent motion will occur in the same plane, since (\ref{eq_GeodSym}) always implies $\dot u^\theta=0$ (or $\dot u^\phi=0$). This is, of course, related to the conservation of angular momentum, as we will learn in the course of this paper.

%%%%%%%%%%%%%%
\subsection{Radial Canonical Quantum Gravity with Static Equal Radius Quantisation}

In order to use the $q$-desics equation (\ref{eq_GeodSym}), one needs the expectation values of the metric operators on the RHS. 
This can in principle be done in
very different approaches to quantum gravity:
Schwarzschild/Kruskal sectors have been quantised using metric and
connection variables
\cite{Kuchar:1994zk,Bergamin:2004us,Ashtekar:2005qt,Gambini:2013ooa,
Modesto:2008im,Gielen:2025ovv}; related constructions were developed in specific QG approaches or for particle-spacetime entanglement~\cite{Pipa:2018bui,Davidson:2012dt,Davidson:2014tda,Kanatchikov:2013xmu,Kanatchikov:2014nua,Kanatchikov:2016neu}.
In this paper we will use 
our complementary Static-Equal-Radius (SER) quantisation of static,
spherically symmetric geometries~\cite{Koch:2025yuz} .
In this section, we briefly recap our results. \\

Let's consider the Einstein Hilbert action with a cosmological constant
\be\label{eq_action0}
S_{GR}=\int d^4x \sqrt{-g}c^3
\frac{R+2\Lambda}{16 \pi G}.
\ee

The line element for spherically symmetric spacetime reads
\be\label{eq_LineElement}
ds^2= n^2(r) g(r) d t^2-\frac{1}{g(r)} dr^2-r^2d\theta^2-r^2\sin^2(\theta)d\phi^2,
\ee
Inserting the line element into the Einstein Hilbert action, integrating out the decoupled coordinates $t$ and the angles, and performing an integration by parts, one arrives at
\be\label{eq_action1}
S =B-\int dr \; n(r) \frac{c^4 Z}{2 G}\left(-1-r^2 \Lambda + g + r \dot g\right) \equiv B - \int dr L(n(r),g(r))
\ee
where $B$ denotes the boundary term coming from integration by parts. We also used the change in notation $\partial_rg \rightarrow \dot g$ and in the present scope of our work, we will discuss properties in the bulk of the theory with $r_i<r<r_f$. In this region, the contribution of the boundary term is just a constant, which can be factored out.\\

We then pass to the Hamiltonian formulation using the Legendre transformation. Since the reduced Lagrangian is singular, the Legendre transformation gives rise to constraints. The corresponding total Hamiltonian is obtained by supplementing the canonical Hamiltonian \(H_c\) with the constraints multiplied by Lagrange multipliers. Therefore, our analysis involves constrained Hamiltonian systems~\cite{Wipf:1993xg,Faddeev:1988qp}
\be
H = H_c + \omega_1 \chi_1 + \omega_2 \chi_2 = p_n \dot n + p \dot g- L + \omega_1 \chi_1 + \omega_2 \chi_2
\ee
\bea
&& H_c = \frac{(1+r^2 (-\Lambda) c^2 -g)p}{r} \quad , \quad p_n=\frac{\partial L}{\partial\dot n}=0 \quad, \quad p=\frac{\partial L}{\partial\dot g}=n(r)\frac{c^5 Z r}{2G}, \\ \nonumber
&& \chi_1= p_n = 0 \quad , \quad \chi_2= \left(\frac{p}{r}-n \Gamma \right) = 0
\eea

When passing to the quantum picture, one realises that the Poisson brackets of the constraints in our framework do not vanish. Such constraints are of second class. As discussed in Ref.~\cite{Dirac:1950pj,dirac2013lectures,Date:2010xr,Matschull:1996up}, for such cases, one usually introduces Dirac brackets and then passes to the quantum picture. 

To prepare quantisation we first have to derive the matrix of the Poisson brackets of the constraints
\be
\{\chi_i,\chi_j \}=P_{ij}.
\ee
We find that
\bea
P_{ij}=\Gamma^2 r^2 \left( 
\begin{array}{cc}
     0&1  \\
    -1 & 0
\end{array}
\right)&.
\eea
To make sure that operators do commute with the constraints $\chi_i$, we define the Dirac brackets
\be
\{ A,B\}^*= \{A,B \}-\{A,\chi_i \}P_{ij}^{-1}\{B,\chi_j \}.
\ee
Now the constraints become strongly zero, which we can use as external identities and not just as results of the equations of motion
\bea\label{eq_Constraint1S}
 p_n &=&0,\\
n &=& \frac{p}{r \Gamma}\label{eq_Constraint2S}.
\eea

This line element  has two unknown functions $n(r)$ and $g(r)$.
In general relativity the solutions for the two functions are
\bea\label{eq_ncl}
n(r)&=&1,\\
\label{eq_gcl}
g(r)&=& 1-  \frac{2 G M}{r}-\frac{r^2 \Lambda }{3}.
\eea

Upon quantisation, the metric functions are promoted to field operators.  
For $g(r)$ we get
\bea
g &\rightarrow &\hat g
\eea
with the canonically conjugated momentum $\hat p$. Both operators $\hat g$ and $\hat p$ obey the canonical commutation relations $ [ \hat g , \hat p ] = i \hbar $ .
The other metric function $g_{00}= n^2 g$ becomes determined through a constraint as
\bea
n&\rightarrow &\hat n
=\frac{\hat p}{r\Gamma}\label{eq_Constraintn},
\eea
where $\Gamma= i  (G/( \hbar))^{1/4} \gamma /(G M_{\langle\hat g  \rangle})$ is a global integration factor with units of energy which we parameterised in terms of the gravitational couplings $G$, the central gravitational mass $M_{\langle\hat g  \rangle}$ and a dimensionless parameter $\gamma$.
The Hamiltonian of this quantum system in the Einstein-Hilbert truncation is given by
\be\label{eq_H}
\hat H=\frac{1-r^2\Lambda }{r}\hat p-\frac{(\hat g \hat p)_W}{r},
\ee
where we remember that observable operator products such as the Hamiltonian are Weyl ordered.
Now having the commutation relation and the Hamiltonian, one can use the Ehrenfest theorem to calculate expectation values. For instance, for the purpose of $q$-desics, we are interested in expectation values of values metric operators composed of product of different powers of operators $\hat g$ and $\hat p$. These can be obtained from the Hamiltonian (\ref{eq_H}) by virtue of the Heisenberg relation 
\be\label{eq_Ehrenfest}
i \hbar\frac{d \langle \hat O\rangle }{dr}= +
\langle [\hat O,\hat H] \rangle +i \hbar \langle \partial_r \hat O\rangle.
\ee
Note further, that in our framework we have rescaled the radius with c such that it has the dimension of time, $r \rightarrow r/c$ and then set $c=1$ for simplicity. Furthermore, due to the choice of our quantisation hypersurface, the variable $r$ is our evolution parameter and plays the same role as time. However, even though this variable then has units of time and one may call it somewhat misleadingly
radial time, it is by no means the “time” which is given from the metric signature, which in the classical context even
changes sign at the crossing of a horizon. Thus, it is the choice of quantisation hypersurface which defines the direction
of the Hamiltonian evolution parameter r and not the special entrance in a given classical space time metric (see Ref.~\cite{Koch:2025yuz} and references therein for more details).
Because of this property of spatial evolution, it has been suggested to us that these spacetime states and their expectation values could also be interpreted as particular {\it quantum reference frames}. Independently how one calls them, they are distinct from a purely classical background and this distinction will e.g. manifest
in the motion of test particles.

%%%%%%%%%%%%%%%%%%%%%%%
%%%%%%%%%%%%%%%%
\section{Symmetries of spacetime with quantized metric} \label{form}
\subsection{Formalism}
%In this section, we introduce the basic differential-geometric structures required for extending spacetime symmetries to an operator-valued metric and to able to later discuss Lie derivatives and killing equation in the presence of operator-valued metric. A more detailed discussion of the underlying manifold, the reduced configuration space, and the associated Hilbert-space structure is provided in Appendix~\ref{sec_app_formalism}.
In this section, we introduce the differential-geometric structures required to formulate spacetime symmetries in the presence of an operator-valued metric. The underlying spacetime manifold and coordinates remain classical, while the metric components are promoted to operators. A more detailed discussion of the manifold, reduced configuration space, and associated Hilbert space is given in Appendix~\ref{sec_app_formalism}.

\subsubsection{Vectors, covectors, and their metric duals}
In our work, we are promoting metric functions to operators but keeping the coordinates $x^{\mu}$ still non quantum. This means that the manifold we have is still smooth which allows us to define coordinates. We therefore treat $ x^\mu $ as ordinary coordinate functions on the classical manifold.
%In differential geometry, the dual to the vector $x^{\mu}$ is the one-form $x_{\mu}$ which is generated by the metric. There, the metric function also serves as a raising and lowering operator or a map for switching between dual these objects. In our formalism, the metric is quantised and we define the following formalism:
%We define raising and lowering of the indices by the metric operator.
%\begin{equation}
%    \hat x_{\mu} =   \hat g_{\mu \nu} x^{\nu} \longrightarrow x_{\mu} = \langle x_{\mu} \rangle =  \langle \hat g_{\mu \nu} \rangle x^{\nu}
%\end{equation}
%Thus, the dual to $x^{\mu}$ is operator-valued due to the nature of metric operator.\\

%{\textcolor{red}{
%??? As discussed today: doing it like this corresponds to the geodescis approximation, not $q$-desics???\\
Let
\be
V=V^\mu\partial_\mu
\ee
be a classical vector field on the underlying spacetime manifold. Since the metric is operator-valued, its metric dual becomes
\be
\hat V_\mu=\hat g_{\mu\nu}V^\nu.
\ee
Taking expectation values gives, for example,
\be
V_\mu \equiv \langle\hat V_\mu\rangle
=
\langle\hat g_{\mu\nu}\rangle V^\nu.
\ee
The coordinate derivatives $\partial_\mu$ remain ordinary derivatives on the classical manifold. Thus, in our work, when we write a covector $V_\mu$, i.e. without $\; \hat{} \;$, we are simply referring to the expectation value $\langle\hat V_\mu\rangle$. 
Similarly, for a classical covector $W_\mu$, one may define
\be
\hat W^\mu=\hat g^{\mu\nu}W_\nu.
\ee

%One can also define the lowering and raising of the indices using the metric operator directly, $ \hat g_{\mu \nu}  x^{\nu} $. However, one must be careful that now the one-form $x_{\mu}$ is an operator object due to the presence of the metric operator. Taking the expectation value of this expression results in $ x_{\mu} =  \langle \hat g_{\mu \nu}  x^{\nu} \rangle = \langle \hat g_{\mu \nu} \rangle x^{\nu}  $ where $x_{\mu}$ is to be understood an expectation valued object, i.e. $x_{\mu} = \langle \hat x_{\mu} \rangle = \langle \hat g_{\mu \nu}  x^{\nu} \rangle$.

\subsubsection{Covariant derivatives}
Given a set of basis vectors $\Vec{e}_{\alpha}$, we can write any vector $\Vec{V}$ as
\begin{equation}
    \Vec{V}= V^{\alpha} \, \Vec{e}_{\alpha}
\end{equation}
Taking the derivative
\begin{equation}
    \frac{\partial \Vec{V}}{\partial x^{\beta}}= \frac{\partial \left( V^{\alpha} \, \Vec{e}_{\alpha} \right) }{\partial x^{\beta}} = \frac{\partial V^{\alpha}}{\partial x^{\beta}} \, \Vec{e}_{\alpha} + V^{\alpha} \, \frac{\partial \Vec{e}_{\alpha}}{\partial x^{\beta}}
\end{equation}
The derivative in the second term is again a vector and one usually writes it as a linear combination of basis vectors
\begin{equation}
    \label{eq:def_Gamma_v1}
    \frac{\partial \Vec{e}_{\alpha}}{\partial x^{\beta}} = \Gamma^{\mu}_{\alpha \beta} \Vec{e}_{\mu}
\end{equation}
Thus, one can write
\begin{equation}
    \label{eq:def_covdev_v1}
    \frac{\partial \Vec{V}}{\partial x^{\beta}} = \frac{\partial V^{\alpha}}{\partial x^{\beta}} \, \Vec{e}_{\alpha} + V^{\alpha} \, \Gamma^{\mu}_{\alpha \beta} \Vec{e}_{\mu} = \left( \frac{\partial V^{\alpha}}{\partial x^{\beta}}+ V^{\mu} \, \Gamma^{\alpha}_{\mu \beta} \right) \Vec{e}_{\alpha} = \nabla_{\beta} V^{\alpha} \, \Vec{e}_{\alpha}
\end{equation}
In our formalism, we will use the covariant derivative as well. The covariant derivative contains the metric $g_{\mu \nu}$ due to the presence of the Christoffel symbols. Having the metrics being promoted to operators and thus the Christoffel symbols being objects of operator nature $\Gamma^{\alpha}_{\mu \nu} \rightarrow \hat \Gamma^{\alpha}_{\mu \nu} $, results in having the covariant derivative also in operator form $ \nabla_{\mu} \rightarrow \hat \nabla_{\mu}$. We define the action of the covariant derivative on some tensor operator as
%Now one can ask, how does this act and how one should use it? The covariant derivative is evaluated by sandwiching it with wave functional $| \Psi(g) \rangle$. The l.h.s. of (\ref{eq:def_Gamma_v1}) is not an operator object and therefore, the r.h.s. should not be of operator nature. We interpret the christoffel symbols which appears on the r.h.s of the equation as the expectation value of the operator form of the christoffel symbol, $\Gamma^{\alpha}_{\mu \nu} = \langle \hat \Gamma^{\alpha}_{\mu \nu} \rangle$. Thus, the action of the covariant derivative is defined through its VEV $\langle \hat \nabla_{\mu} \rangle$. Hence,
\begin{equation}
   \langle \hat \nabla_{\mu} V^{\nu} \rangle  = \partial_{\mu} V^{\nu} + \langle \hat \Gamma_{\mu \rho}^{\nu} \rangle V^{\rho}
\end{equation}
\begin{equation}
     \langle \hat \nabla_{\rho} \hat O_{\mu \nu} \rangle = \langle  \partial_{\rho}\hat O_{\mu \nu} - \hat O_{\nu \alpha} \hat \Gamma^{\alpha}_{\rho \mu} - \hat O_{\mu \alpha} \hat \Gamma^{\alpha}_{\rho \nu}  \rangle
\end{equation}

With the operator form of the Christoffel symbol given as (see Ref.~\cite{Koch:2025qzv})
\begin{equation}
    \hat \Gamma^{\alpha}_{\mu \nu} = \frac{1}{2} \hat g^{\alpha \beta} \left( \hat g_{\mu \beta , \nu} + \hat g_{\nu \beta , \mu} - \hat g_{\mu \nu , \beta} \right),
\end{equation}
one immediately notices that the metricity condition is satisfied at the operator-valued level for the case where Christophel symbols are symmetric in the lower indices and using $\hat g_{\mu \alpha} \hat g^{\alpha \nu} = \delta^{\nu}_{\mu} $, where $\hat g^{\alpha \nu} $ is the inverse of the metric operator. Then we have
\begin{equation}
     \langle \hat \nabla_{\rho} \hat g_{\mu \nu} \rangle = \langle  \partial_{\rho}\hat g_{\mu \nu} - \hat g_{\nu \alpha} \hat \Gamma^{\alpha}_{\rho \mu} - \hat g_{\mu \alpha} \hat \Gamma^{\alpha}_{\rho \nu}  \rangle =  \left\langle  \partial_{\rho} \hat g_{\mu \nu} - \hat g_{\nu \alpha}  \frac{1}{2} \hat g^{\alpha \beta} \left( \hat g_{\mu \beta , \rho} + \hat g_{\rho \beta , \mu} - \hat g_{\mu \rho , \beta} \right) - \hat g_{\mu \alpha} \frac{1}{2} \hat g^{\alpha \beta} \left( \hat g_{\rho \beta , \nu} + \hat g_{\nu \beta , \rho} - \hat g_{\rho \nu , \beta} \right) \right\rangle = 0.
\end{equation}
Thus, within the above inverse-operator and ordering prescription, the operator metricity condition is satisfied. This statement concerns metric compatibility of the operator pair $\left( \hat g_{\mu \nu}, \hat \Gamma^{\alpha}_{\mu \nu} \right) $. It does not imply that $ \langle \hat \Gamma^{\alpha}_{\mu \nu} \rangle $ is the Levi-Civita connection of the expectation-value metric $ \langle \hat g_{\mu \nu} \rangle$. Furthermore, it should be mentioned that we use the Weyl ordering for the product of operators inside the expectation values. The operator-valued level metricity condition also allows us to write
\begin{equation}
    \langle \hat \nabla_{\alpha} \hat O_{\beta} \rangle = \langle \hat \nabla_{\alpha} \hat g_{\beta \sigma} \hat O^{\sigma} \rangle = \langle \hat g_{\beta \sigma} \hat \nabla_{\alpha}  \hat O^{\sigma} \rangle = \left\langle \hat g_{\beta \sigma} \left( \hat \partial_{\alpha}  \hat O^{\sigma} + \hat \Gamma^{\sigma}_{\alpha \rho} \hat O^{\rho}\right) \right\rangle  = \left\langle \hat g_{\beta \sigma} \hat \partial_{\alpha}  \hat O^{\sigma} \right\rangle + \left\langle \hat g_{\beta \sigma}  \hat \Gamma^{\sigma}_{\alpha \rho} \hat O^{\rho} \right\rangle  
\end{equation}
Also the covariant derivative itself with upper index can be written as $\hat \nabla^{\alpha}= \hat g^{\alpha \beta }  \hat \nabla_{\beta}$.
%The covariant derivative act on the expectation value of the metric operator. We can define the metricity by having $\nabla_{\rho} g_{\mu \nu}  = 0 $. Keep in mind that in our formalism, we have $\langle \hat \Gamma^{\alpha}_{\mu \nu} (\hat g) \rangle$ which is different than $\Gamma^{\alpha}_{\rho \nu} ( \langle \hat g \rangle)$.
It is worth to distinguish this argumentation from the previous arguments on non-metricity theories. % The metricity or the vanishing of the covariant derivative of the metric is satisfied in our framework at the operator level, where the q-desics equation itself is analouge to theories which have non-metricity features at the classical level.
Within the inverse-operator and ordering prescription adopted here, metricity is satisfied at the operator level. This should, however, be distinguished from metric compatibility at the level of expectation values. In general, the connection $\langle \hat{\Gamma}^{\alpha}_{\mu\nu}\rangle$ is different from the connection with the expectation values of metric $\langle \hat g_{\mu\nu}\rangle$, i.e. $\langle \hat{\Gamma}^{\alpha}_{\mu\nu} \left( \hat g \right) \rangle \neq \Gamma^{\alpha}_{\mu\nu} \left( \langle \hat g \rangle \right)$ and the difference is encoded in the $\Delta^\alpha_{\mu \nu}$ as expressed in eq.~\ref{eq_Covariance}. Consequently, the q-desic equations may exhibit some effective non-metricity features despite metricity being imposed at the operator level. These contributions originate from the fact that product of expectation values is not the same as expectation value of product of operators, i.e. $\langle \hat A \hat B\rangle \neq \langle \hat A\rangle\langle \hat B\rangle$, and the non-metricity behaviour of the $q-$desics equation comes from the difference of these as discussed in section~\ref{sec_qdesics}.

\subsection{Symmetries and metric operators}

We now extend the classical description of spacetime symmetries to the case in which the metric components are promoted to operators. Since the spacetime coordinates and the underlying manifold remain classical, infinitesimal coordinate transformations can be treated in the usual manner, while their action on the geometry must be formulated at the operator level. This leads naturally to an operator-valued Lie derivative and, subsequently, to a generalised Killing condition for the quantised metric. In Ref.~\cite{Koch:2025qzv}, the quantum version of the geodesics equation in such spacetime has been derived. 

\subsubsection{Lie Derivative}
We wish to derive a similar equation for the Lie derivative in a quantised spacetime where the metric tensor is an operator $\hat g_{\mu \nu}$. Let $x^{' \rho}  =  x^{\rho} +  \epsilon\, \xi^{\rho}$
\begin{equation}
    \hat g^{'}_{\mu \nu} (x') = \frac{\partial x^{\alpha}}{\partial x^{' \mu}} \hat g_{\alpha \beta} (x) \frac{\partial x^{\beta}}{\partial x^{' \nu}} = \left( \delta_{\mu}^{\alpha} - \epsilon \partial_{\mu} \xi^{\alpha} \right) \hat g_{\alpha \beta} \left( \delta_{\nu}^{\beta} - \epsilon\partial_{\nu} \xi^{\beta} \right) 
\end{equation}
\begin{equation}
    \hat g^{'}_{\mu \nu} (x^{' \alpha}) \approx \delta_{\mu}^{\alpha} \hat g_{\alpha \beta} \delta_{\nu}^{\beta} - \epsilon \left( \delta_{\mu}^{\alpha} \hat g_{\alpha \beta} \partial_{\nu} \xi^{\beta} +  \delta_{\nu}^{\beta} \hat g_{\alpha \beta}  \partial_{\mu} \xi^{\alpha} \right) = \hat g_{\mu \nu} (x^{\alpha}) \approx \hat g_{\mu \nu} (x) -  \epsilon \left( \hat g_{\mu \beta} \partial_{\nu} \xi^{\beta} + \hat g_{\alpha \nu}  \partial_{\mu} \xi^{\alpha} \right)
\end{equation}
On the other hand, we can expand $\hat g_{\mu \nu} (x')$ up to first order
\begin{equation}
    \hat g_{\mu \nu} (x^{' \rho} ) = \hat g_{\mu \nu} (x^{\rho} + \xi^{\rho} \epsilon ) \approx  \hat g_{\mu \nu} (x ) +  \xi^{\rho}  \partial_{\rho} \hat g_{\mu \nu} (x) \epsilon
\end{equation}
Taking the difference of the two equations and then the limit $\epsilon \rightarrow 0$
\begin{equation}
    \lim_{\epsilon \to 0} \frac{\hat g_{\mu \nu} (x^{' \rho} ) - \hat g^{'}_{\mu \nu} (x^{' \alpha})  }{\epsilon} = \xi^{\rho}  \partial_{\rho} \hat g_{\mu \nu}   + \hat g_{\mu \beta} \partial_{\nu} \xi^{\beta} + \hat g_{\alpha \nu}  \partial_{\mu} \xi^{\alpha} \equiv \left( \hat {\mathcal{L}_{\xi} }\hat g \right)_{\mu \nu}
\end{equation}
%using $\partial_{\mu} u^{\alpha} = \hat \nabla_{\mu} u^{\alpha} - \hat \Gamma_{\mu \beta}^{\alpha} u^{\beta}$ and $\partial_{\rho} \hat g_{\mu \nu} = \hat \nabla{\rho} \hat g_{\mu \nu} + \hat \Gamma^{\alpha}_{\rho \mu} \hat g_{\nu \alpha} + \hat \Gamma^{\alpha}_{\rho \nu} \hat g_{\mu \alpha}$ we can write the Lie derivative for the metric operators 
%\newline
%The only non-vanishing Christoffel symbols are $\hat \Gamma^{0}_{10}$, $\hat \Gamma^{0}_{01}$, $\hat \Gamma^{1}_{00}$, $\hat \Gamma^{1}_{11}$, $\hat \Gamma^{1}_{22}$, $\hat \Gamma^{1}_{33}$, $\hat \Gamma^{3}_{23}$, $\hat \Gamma^{3}_{32}$, $\hat \Gamma^{2}_{12}$, $\hat \Gamma^{2}_{21}$, $\hat \Gamma^{2}_{33}$, $\hat \Gamma^{3}_{13}$, $\hat \Gamma^{3}_{31}$.\\
%\newline

Before taking the expectation value, we use $\partial_{\mu} \xi^{\alpha} = \hat \nabla_{\mu} \xi^{\alpha} - \hat \Gamma_{\mu \beta}^{\alpha} \xi^{\beta}$,  $\partial_{\rho} \hat g_{\mu \nu} = \hat \nabla_{\rho}  \hat g_{\mu \nu}+ \hat g_{\nu \alpha} \hat \Gamma^{\alpha}_{\rho \mu}  +  \hat g_{\mu \alpha} \hat \Gamma^{\alpha}_{\rho \nu}  $ and $  \hat \xi_{\mu} =  \hat g_{\mu \nu} \xi^{\nu} $  to write the Lie derivative for the metric operators 
\begin{equation}
    \left( \mathcal{L}_{\xi} \hat g \right)_{\mu \nu} \equiv \hat \nabla_{\mu} \hat \xi_{\nu} + \hat \nabla_{\nu} \hat \xi_{\mu}
\end{equation}
Taking the expectation value, we have
\begin{equation}
    \left\langle \left( \mathcal{L}_{\xi} \hat g \right)_{\mu \nu} \right\rangle \equiv \langle \hat \nabla_{\mu} \hat \xi_{\nu} + \hat \nabla_{\nu} \hat \xi_{\mu} \rangle = \langle \hat \nabla_{\mu} \hat \xi_{\nu} \rangle + \langle \hat \nabla_{\nu} \hat \xi_{\mu} \rangle
\end{equation}

%%%%%%%%%%%%%%%%%%%%%%%%%%%
\subsubsection{$q$-illing Equation}
The vanishing of Lie derivative signals the presence of symmetries. $q$-illing vectors are those vectors $\xi^{\mu}$ which make the Lie derivative vanish, namely
%\begin{equation}\label{eq_LieDer1}
%      \left( \hat {\mathcal{L}_{\xi} }\hat g \right)_{\mu \nu} =  \xi^{\rho}  \partial_{\rho} \hat g_{\mu \nu}   + \hat g_{\mu \beta} \partial_{\nu} \xi^{\beta} + \hat g_{\alpha \nu}  \partial_{\mu} \xi^{\alpha} = 0

%%\quad \text{where} \quad
%    \xi^{\mu} = \xi^{0}\partial_0 + \xi^{1}\partial_1 + \xi^{2}\partial_2 + \xi^{3}\partial_3, % \quad \text{and} \quad \xi_{\mu} =  \langle \hat g_{\mu \nu} \rangle \xi^{\nu}
%\end{equation}
%or equivalently 
\begin{equation}\label{eq_Killing2}
    \hat \nabla_{\mu} \hat \xi_{\nu} + \hat \nabla_{\nu} \hat \xi_{\mu}  = 0. %\quad \text{or} \quad \hat \nabla_{\mu} \hat \xi_{\nu} + \hat \nabla_{\nu} \hat \xi_{\mu}  = 0 
\end{equation}

This equation is the analog of the killing equation in the presence of metric operators instead of metric functions. We call this equation {\it $q$-illing Equation }.\\

To obtain a tractable effective form of the operator $q$-illing 
equation (\ref{eq_Killing2}), we introduce the ansatz
\be \label{eq_effective_qilling_ansatz}
\hat \xi_\mu \approx \zeta_\mu \hat 1,
\ee
is an ordinary vector field on the underlying classical manifold. Inserting this ansatz in the $q$-illing equation and taking the expectation value gives
\be\label{eq_Qilling}
%\langle \hat \nabla_{\mu} \hat \zeta_{\nu} + \hat \nabla_{\nu} \hat \zeta_{\mu} \rangle 
%=
  \partial_{\mu} \zeta_{\nu} -
\langle\hat \Gamma^\alpha_{\mu \nu} \rangle\zeta_\alpha
+ \partial_{\nu} \zeta_{\mu}  -
\langle\hat \Gamma^\alpha_{\nu \mu} \rangle \zeta_\alpha=0
\ee
We refer to solutions $\zeta_\mu$ of equation~(\ref{eq_Qilling}) as effective $q$-illing covectors. They represent the operator q-illing condition within the approximation $\hat \xi_\mu \approx \zeta_\mu \hat 1$. In general, equation~(\ref{eq_Qilling}) is not identical to the ordinary Killing equation constructed from the expectation-value metric $\langle\hat g_{\mu\nu}\rangle $, since $\left\langle 
\hat \Gamma^\mu_{\alpha \beta}
\right\rangle \neq 
 \Gamma^\mu_{\alpha \beta}$. However, depending on the quantum properties of the underlying spacetime, the effective $q$-illing covectors $\zeta_\mu$ may coincide with the effective $q$-illing covectors of the averaged space-time background appearing in general relativity.

%This equation has the same structure as in the classical case and the vacuum expectation value of the metric operators gives the classical results. For the spherically symmetric static spacetime with metric operators, the killing vectors are the same as one would have in the classical limit.

%%%%%%%%%%%%%%%%%%%%
\section{Conserved quantities} \label{conserved}

Before we turn to the concrete discussion on the conservation laws for $q$-desic motion, it is instructive to start with a more conceptual argument.

%%%%%%%%%%%%%
\subsection{The order of things in the $q$-desic context}

At the current stage, the implementation of $q$-desics in concrete examples requires several 
conceptual and practical  steps. These steps are
\begin{itemize}
    \item[$\delta_x^I$] Variation with respect to the particle degrees of freedom and isolation of the 
    	acceleration times the spacetime unity operator $\hat 1$
    \item[$\hat g^{-1}$] Isolate the acceleration times the unit operator $\hat 1$, if necessary by applying inverse metric operators $\hat g^{-1}$    
    \item[$\langle.. \rangle $] Take expectation value of the metric operators
    \item[${\mathcal{S}}_{}$] Assume a symmetry when evaluating the above expectation values ${\mathcal{S}}_{\langle . \rangle}$
    and when evaluating the $q$-desic equation ${\mathcal{S}}_{\ddot x}$
\end{itemize}
While the steps $\delta_x^I$, $\hat g^{-1}$, $\langle.. \rangle $ are conceptual for the $q$-desic approach, the last step is practical in the sense 
that it simplifies the
calculations, but they are only applicable to very specific systems.
The new aspect of the $q$-desic is to highlight the importance of the order in which these operations are done.
I particular, if one takes $\langle.. \rangle $ too early, one looses information and falls back from the $q$-desic back
to the classical geodesics.

%\begin{figure}
%    \centering
%    \includegraphics[width=0.8\linewidth]{figures/order_of_operations_creative_highres.pdf}
%    \caption{Caption}
%    \label{fig:placeholder}
%\end{figure}

When we derived the $q$-desic equation we learned that operation $\delta_x^I$ of varying an action and isolating the acceleration does not commute with taking the EV of the metric degrees of freedom $\langle \hat O\rangle$.
Thus, if we consider in a qualitative argument the information content ${\mathcal{I}}$ of these two procedures, the $q$-desics contain typically more information than the geodesics
\be\label{eq_inequal1}
{\mathcal{I}}(\langle \delta_x^I \hat {\mathcal{L}}\rangle)\ge {\mathcal{I}}(\delta_x^I\langle  \hat {\mathcal{L}}\rangle).
\ee
For calculating concrete examples of metric EVs we have to rely on symmetries $\mathcal{S}$, such as for example assuming homogeneity or spherical symmetry. 
When information content is concerned, symmetry conditions $\mathcal{S}$ and other operations $\hat O$, such as taking EVs do not commute either and it is typically true that 
\be\label{eq_inequal2}
{\mathcal{I}}(\mathcal{S} \hat O \hat {\mathcal{L}})
\ge
{\mathcal{I}}( \hat O\mathcal{S}\hat {\mathcal{L}}).
\ee
For this reason, it is generally believed that while minisuperspace models can provide some information about quantum gravity, they still may contain less information than the corresponding  full quantum gravitational theory, where the symmetry is imposed a posteriori. 
By combining (\ref{eq_inequal1}) and (\ref{eq_inequal2}) it becomes clear, that for the same reason
\be\label{eq_inequal3}
{\mathcal{I}}({\mathcal{S}}\langle \delta_x^I \hat {\mathcal{L}}\rangle)\ge {\mathcal{I}}(\delta_x^I\langle  {\mathcal{S}}\hat {\mathcal{L}}\rangle).
\ee
Thus, in the context of $q$-desics, we expect that it  makes a difference whether a symmetry is imposed a posteriori or assumed a priori.

Two symmetries that are particularly important in the context of geodesics are time and rotational invariance, leading to the conserved quantities Energy $E$ and angular  momentum $l$.
These relations have to be re-evaluated in the context of  $q$-desics. Given the inequality (\ref{eq_inequal3}), 
we should do this a posteriori, by contemplating the space of solutions of these equations.

%%%%%%%%%%%%%%%%%%%
\subsection{4-velocity conservation and mass definition}

To address the conservation of $u^\mu u_\mu$, we need to delay the gauge fixing in the $q$-desic equation, which was given in (\ref{eq_qdesic}).
It is now useful to use our definition of the connection-covariance 
\be\label{eq_defDelta}
\Delta^\mu_{\alpha \beta}=
\left \langle \hat \Gamma^\mu_{\alpha \beta}\right\rangle-  \Gamma^\mu_{\alpha \beta}(\langle \hat g_{..}\rangle).
\ee
With this, we can express the $q$-desic equation in terms of the Levi-Civita connection calculated from the metric EVs and the covariance of the space-time state
\be\label{eq_Covqdesic}
\frac{d u^\mu}{d\lambda}=
-\Gamma^\mu_{\alpha \beta}(\langle \hat g \rangle) u^\alpha u^\beta
-\Delta^\mu_{\alpha \beta}u^\alpha u^\beta
- \frac{\dot \omega}{\omega}u^\mu.
\ee
We realise, that the quantum corrections introduce modifications of the classical geodesic equations, which are exactly of the auto-parallel type,
where one has a difference between the connection and the Levi-Civita symbol $\left \{ ^\mu_{\; \alpha \beta}\right\}$.

To check whether a worldline, which initially has $u^\mu u_\mu=k^2$ (with $k^2=1$ fore timelike, $k^2=0$ for light-like, and $k^2=-1$ for space-like motion), will conserve it's characteristics.
Thus, we examine the contraction $\langle u^\mu \hat g_{\mu \nu} u^\nu \rangle=u^\mu  \langle \hat g_{\mu \nu}\rangle u^\nu =u^2$, while using the $q$-desic equation in its form (\ref{eq_Covqdesic}). We find for the derivative that
\be\label{eq_v2NonConserved}
\frac{d}{d \lambda} \left( u^\mu \langle\hat g_{\mu \nu}\rangle u^\nu\right)
= -2 u^\alpha u^\beta u_\mu \Delta ^\mu_{\alpha \beta} - 2 \frac{\dot \omega}{\omega} 
\left( u^\mu \langle\hat g_{\mu \nu}\rangle u^\nu\right).
\ee

However, one can choose a gauge such that this particular four velocity (subject to the definition of $\lambda$) is conserved
\bea
\left.\frac{\dot \omega}{\omega} \right|_q &=& -\frac{u^\alpha u^\beta u_\mu \Delta ^\mu_{\alpha \beta}}{u^2}.\label{eq_gaugefix2}
\eea
The gauge choice (\ref{eq_gaugefix2}), which makes $q$-desic 4-velocities squares conserved.
The 
geometrical interpretation is obtained 
from a normalisation condition
\be\label{eq_Unorm}
U^\mu = \frac{u^\mu}{\sqrt{u^\alpha \langle \hat g_{\alpha \beta} \rangle u^\beta}}\equiv \frac{dx^\mu}{d\tau},
\ee
where we introduced the infinitesimal proper time parameter 
\be\label{eq_ProperTime}
d\tau = d\lambda
\sqrt{u^\alpha \langle \hat g_{\alpha \beta} \rangle u^\beta} .
\ee
Now, it is straightforward to explicitly confirm that
\be\label{eq_U2conserved}
\frac{d}{d\lambda}\left( U^\mu  \langle \hat g_{\mu \nu}\rangle U^\nu\right)=0.
\ee
We may use the normalised four velocity (\ref{eq_Unorm}) to define a momentum in proper time
\be \label{eq_defp}
p_{PT}^\mu = m_{PT}  U^\mu, \quad {\text{with}} \quad
m_{PT}^2=p_{PT}^\mu p^{PT}_\mu 
\ee
where $m_{PT}$ is the mass in the proper time frame.
This mass does not change along the trajectory and is thus a constant of motion
\be
\frac{d}{d\lambda} m_{PT}=0.
\ee
However, the above $m_{PT}$ is different from the usual definition of mass, which reads
\be\label{eq_defmUsual}
m^2=p^\mu \langle \hat g_{\mu \nu}\rangle p^\nu.
\ee
This definition is  based on the momentum
\be \label{eq_defpUsual}
p^\mu = \omega u^\mu.
\ee
The mass (\ref{eq_defmUsual}) has the advantage that it is manifestly gauge invariant, but it has the disadvantage that it is not locally conserved along a $q$-desic trajectory. Due to the non-metricity (\ref{eq_defDelta}) one gets
\be
\frac{d}{d\lambda} m^2= -p^\mu p^\nu p_\alpha \Delta^{\alpha}_{\mu \nu}.
\ee
Thus, for a mass defined by (\ref{eq_defmUsual}), quantum gravitational effects can accumulate a ``mass memory effect'' along a $q$-desic path.
To settle the discussion on the mass definition, it will be essential to clarify, which quantities we actually measure in which sort of experiment and then draw the connection to candidate definitions such as the conserved (\ref{eq_defp}) or the gauge invariant (\ref{eq_defpUsual}).

Note that for null velocities $u^\alpha \langle \hat g_{\alpha \beta}\rangle u^\beta=0$ the definition of a proper time is not possible. We will comment on this in the discussion section.

%%%%%%%%%%%%%%%%%%%
\subsection{Formal definition of conserved quantities}

Given a classical Killing vector (\ref{eq_Killing}), the associated  
quantity (\ref{eq_conserved1}) is only conserved if the test particle moves along a geodesic.
However, for a test particle traveling on a QG background both the Killing vectors~(\ref{eq_Killing}) and the evolution equation  (\ref{eq_qdesic})
change and thus, the definition
(\ref{eq_conserved1}) will no longer provide a conserved quantity any more.
The most straightforward attempt to generalise the conserved quantities (\ref{eq_conserved1}) to a QG background is 
to replace Killing vectors by the effective $q$-illing covectors $\xi_\mu^i\rightarrow \zeta_\mu^i$ and
geodesic motion by $q$-desic motion
\be\label{eq_conservedq}
C_q^i=  \sigma_i \omega  \zeta_\mu^i u^\mu.
\ee
To check whether this quantity is conserved in a QG background, let's calculate the derivative of (\ref{eq_conservedq}) explicitly
\bea
\frac{d C_q^i}{d\lambda}&=&\frac{d}{d\lambda}
\left( \sigma_i \omega\, \zeta^i_\mu u^\mu\right)\\ 
\nonumber&=&
\sigma_i\left( \left(\frac{d}{d\lambda} \omega \right)\zeta_\mu^i u^\mu+
\omega \left( \frac{d}{d\lambda }\zeta_\mu^i \right) u^\mu +\omega  \zeta_\mu^i \left( \frac{d}{d\lambda}u^\mu \right) \right)\\ \nonumber
&=& 0,
\eea
where we used $\frac{d}{d\lambda} = \frac{d x^\nu}{d\lambda}\frac{d}{dx^\nu}= u^\nu \frac{d}{dx^\nu}$, the $q$-illing relation (\ref{eq_Qilling}) for $\dot \zeta_\mu= \partial_\nu \zeta_\mu u^\nu$, and the $q$-desic equation (\ref{eq_qdesic}) for the term $\frac{d}{d\lambda} u^\mu$ .
This confirms that (\ref{eq_conservedq}) is indeed the correct definition of a conserved quantity in the context of $q$-desic motion.

%%%%%%%%%%%%%%%%%%%%%%%%%%%%%%
\section{Case study for static spherically symmetric quantum spacetime} \label{caseofstudy}

We can now explore the, up to now developed, concepts for the concrete example of static spherically symmetric quantum spacetime.

%%%%%%%%%%%%%%%%%%%
\subsection{$q$-illing condition}
\label{sec_$q$-illing}

To find the conserved quantities from the symmetries of our spacetime, one needs to solve eq.~\ref{eq_Qilling}. Then, one can use the solutions and contract them with the corresponding component of the four-velocity of the particle on its trajectory along the $q$-desics. Doing so, one is able to have the conserved quantities of the particle on $q$-desics. Eq.~\ref{eq_Qilling} gives ten differential equations for the components of the effective q-illing covector $\zeta_{\mu}$. For the detailed derivation, we refer the reader to appendix~\ref{appendix_qillingvectors}. Having eq.~\ref{eq_Qilling} solved, one finds that the $\zeta_t$, $\zeta_r$, $\zeta_\theta $ and $\zeta_\phi$ are given by
\be
\label{sol_zeta0_v1}
\zeta_t = c_0 e^{\int dr A(r)} \quad \text{with}\quad A(r) = \frac{1}{r} \left(  \langle \hat g^{-1} \rangle \left( 1-r^2\Lambda \right)  -1 \right)
\ee

\be
\label{sol_zeta1_v1}
\zeta_r =0
\ee

\be
\label{sol_zeta2_v1}
\zeta_\theta =r^2 \left( c_1 \cos (\phi) + c_2 \sin( \phi) \right)
\ee

\be
\label{sol_zeta3_v1}
\zeta_\phi =c_3 r^2 \sin^2 (\theta) - r^2 \cos(\theta) \sin( \theta) \left( c_1 \sin (\phi) - c_2 \cos( \phi) \right)
\ee

%%%%%%%%%%%%%%%%%%%
\subsection{Conserved quantities from effective $q$-illing covectors}
In the following section, we determine the conserved quantities, energy and angular momentum, of the particle along $q$-desics using effective $q$-illing covectors.
\subsubsection{Energy}
The energy of the system is given by (\ref{eq_conservedq} and \ref{sol_zeta0_v1}) with $i=0$ for the former. 
\be \label{eq_energy_qilling_v1}
E=  \sigma_0 \omega  \zeta_t u^t =\sigma_0\omega c_0 e^{\int dr A(r)} u^t =  u^t \omega e^{\int dr A(r)}
\ee
where we have imposed $\sigma_0 c_0 = 1$, since both $\sigma_0$ and $c_0$ are just constants. One can check that eq.~\ref{eq_energy_qilling_v1} recovers the classical energy in general relativity for the case with $\Lambda =0$ and in the limit $\langle \hat g^{-1} \rangle \rightarrow 1/g $, i.e. classical limit.

Now we would like to solve eq.~\ref{eq_energy_qilling_v1} for the quantum gravity case. Our first challenge is to find an analytical expression for the expectation value of the inverse metric operator. 
This can be achieved in different ways, as discussed in the appendix~\ref{sec_divergencies}.
Here, we will continue with the {\it weak field expansion}, which was already outlined in~\cite{Koch:2025qzv}. It allows us to expand $\langle \hat g^{-1} \rangle$ in terms of the expectation value of the metric itself and its higher powers (\ref{eq_ginvEV}).
The integral $\int dr A(r)$ can then be written approximately as
\bea
\int dr A(r) &=& \int dr  \frac{1}{r} \left(  \left\langle \frac{1}{\hat g} \right\rangle \left( 1-r^2\Lambda \right)  -1 \right) = \int dr  \frac{1}{r} \left(  \sum_{n=1}^\infty \sum_{l=0}^{n-1}  \left( \begin{array}{c}
         n-1  \\
         l 
    \end{array}\right)
      (-1)^{l} \langle \hat g^l \rangle \left( 1 -r^2 \Lambda \right) -\frac{1}{r} \right) \\ \nonumber  & \approx &  \int dr  \frac{1}{r}\left(  \left(3 - 3\langle \hat g \rangle + \langle \hat g^2 \rangle\right) \left(1  -r^2 \Lambda \right) -1 \right) + ... 
       \\ \nonumber & \approx &  \left(\frac{2}{9} \Lambda ^2 r^3 C_{1,0}-\frac{ C_{2,0}}{2 r^2}+\frac{1}{3} \Lambda  r C_{1,0}- \Lambda  \log (r/r_0) C_{2,0}+\frac{ C_{1,0}}{r}-\frac{1}{54} \Lambda ^3 r^6-\frac{\Lambda ^2 r^4}{18}-\frac{1}{3} \Lambda  r^2\right),
\eea
where $r_0$ is an initial value for $r$, and $C_{1,0}$ and $C_{2,0}$ are constants discussed in Ref.~\cite{Koch:2025yuz}. Expanding the expansion of the exponential function $\left( e^x \approx 1 + x + ... \right)$ we can write $e^{\int dr A(r)}$ as
\be
 e^{\int dr A(r)} \approx \left( 1 +  \frac{C_{1,0}}{r} - \frac{\Lambda r^2}{3} - \frac{C_{2,0}}{2r^2} + \frac{2C_{1,0}\Lambda^2 r^3}{9} + \frac{C_{1,0}\Lambda r}{3} - \frac{\Lambda^2 r^4}{18} + \frac{\Lambda^3 r^6}{54} - \Lambda  \log (r/r_0) C_{2,0} \right).
\ee
The first three terms in the above equation are just the classical solution of the metric function. The energy E is given by 
\bea
E & \approx & u^t \omega   \left( 1 +  \frac{C_{1,0}}{r} - \frac{\Lambda r^2}{3} - \frac{C_{2,0}}{2r^2} + \frac{2C_{1,0}\Lambda^2 r^3}{9} + \frac{C_{1,0}\Lambda r}{3} - \frac{\Lambda^2 r^4}{18} + \frac{\Lambda^3 r^6}{54} - \Lambda  \log (r/r_0) C_{2,0}  \right) \\ \nonumber & = &   u^t \omega \left( 1 +  \frac{C_{1,0}}{r} - \frac{\Lambda r^2}{3} \right) + u^t \omega \left(- \frac{C_{2,0}}{2r^2} + \frac{2C_{1,0}\Lambda^2 r^3}{9} + \frac{C_{1,0}\Lambda r}{3} - \frac{\Lambda^2 r^4}{18} + \frac{\Lambda^3 r^6}{54} - \Lambda  \log (r/r_0) C_{2,0} \right).
\eea
The first term corresponds to the energy obtained in the classical case while the second term could be interpreted as the first order correction term to the energy of the particle on its trajectory. In our expression, we have constants $C_{1,0}$ and  $C_{2,0}$ which appear in calculating the inverse of the metric operator $\hat g^{-1}$ coming from expectation values of higher powers of $\hat g$ in our expansion.
We can fix these constant by setting $\Lambda =0$, and comparing the expansion of $\langle \hat g^{-1} \rangle$ with the expansion of the inverse metric in classical limit. 
\be
\left\langle \frac{1}{\hat g} \right\rangle \approx \left(3 - 3\langle \hat g \rangle + \langle \hat g^2 \rangle\right) = 1- \frac{C_{1,0}}{r} + \frac{C_{2,0}}{r^2}
\ee
and 
\be
\frac{1}{1 - \frac{2GM}{r}} \approx 1 +  \frac{2GM}{r} + \frac{4 G^2 M^2}{r^2}.
\ee
Thus, by comparing the classical limit, we can set $C_{1,0} = -2GM$ and $C_{2,0}= 4G^2 M^2$ to recover the classical solution.
Moreover, for the constant $C_{1,0}$, one can compare the expectation value of metric operator $\hat g$ with the classical solution for metric function $g$. Doing so, one can fix the constant $C_{1,0}=-2GM$ in order to recover the classical limit.

%??? I think now, that these inverse operators are very dangerous, 
%possible solution we should discuss: Multiply the $q$-desic operator equation by $\hat g$ before Weyl ordering and before EV. 

%%%%%%%%%%%%%%%%%%%%%%%
\subsubsection{Angular Momentum}
The three components of angular momentum without the $\omega$-term are given by
\be \label{eq_Lx}
L_x = - r^2 \left( \sin ( \phi) \dot \theta + \sin ( \theta) \cos (\theta ) \cos ( \phi ) \dot \phi \right)  
\ee
\be \label{eq_Ly} 
L_y = r^2 \left( \cos ( \phi) \dot \theta - \sin ( \theta) \cos (\theta ) \sin ( \phi ) \dot \phi \right)  
\ee
\be \label{eq_Lz}
L_z = r^2 \sin^2 ( \theta ) \dot \phi
\ee
In the following, we show by using $q$-illing covectors shows that these three are actually conserved along a geodesics. Our effective $q$-illing covectors has four component $\zeta_{\mu}= \left( \zeta_t,\zeta_r, \zeta_{\theta} ,\zeta_{\phi} \right)$. In section~\ref{sec_$q$-illing}, we found that the radial component of the effective $q$-illing covectors $\zeta_r$ is zero. The first component $\zeta_t$ is the time component and corresponds to the conservation of energy. The other two components form the angular part of the effective $q$-illing covectors and are responsible for the conservation of angular momentum.
\be
C_{angular} = \sigma_{angular}\omega \left( \zeta_{\theta} u^{\theta} + \zeta_{\phi} u^{\phi} \right)
\ee
Inserting the effective $q$-illing vector (\ref{sol_zeta2_v1} and \ref{sol_zeta3_v1}), we obtain
\bea
C_{angular} &=& \sigma_{angular}\omega \left( \left( c_1 r^2 \cos (\phi) + c_2 r^2 \sin( \phi) \right) u^\theta + \left( c_3 r^2 \sin^2 (\theta) - r^2 \cos(\theta) \sin( \theta) \left( c_1 \sin (\phi) - c_2 \cos( \phi) \right) u^{\phi} \right) \right) \\ \nonumber
&=& c_1 \left( \sigma_{angular}\omega \left( r^2 \cos (\phi) u^\theta - r^2 \cos(\theta) \sin( \theta) \sin (\phi) u^\phi \right) \right)
+ c_2 \left( \sigma_{angular}\omega \left( r^2 \sin( \phi) u^\theta + r^2 \cos(\theta) \sin( \theta) \cos (\phi) u^\phi \right) \right) \\ \nonumber 
&+&
c_3 \left( \sigma_{angular}\omega \left( r^2 \sin^2 ( \theta) u^\phi \right) \right)
\eea
where in the final equality, we grouped the coefficients of $c_i$ together. Since $c_i$ are independent constants, this means that if $C_{angular}$ is conserved, then each of the expressions in the paranteses with coefficients $c_1$, $c_2$ and $c_3$ is conserved independently.
%\bea
%\frac{d C_{angular}}{ d\lambda} = 0 & \longrightarrow & \frac{d}{d \lambda} \left( \frac{d_{angular}}{\omega} \left( r^2 \cos (\phi) u^\theta - r^2 \cos(\theta) \sin( \theta) \sin (\phi) u^\phi \right) \right) = 0, 
%\\ \nonumber
%& \longrightarrow & \frac{d}{d \lambda} \left( \frac{d_{angular}}{\omega} \left( r^2 \sin( \phi) u^\theta + r^2 \cos(\theta) \sin( \theta) \cos (\phi) u^\phi \right) \right) = 0, 
%\\ \nonumber
%& \longrightarrow & \frac{d}{d \lambda} \left( \frac{d_{angular}}{\omega} \left( r^2 \sin^2 ( \theta) u^\phi \right) \right) = 0
%\eea
\begin{equation}
\label{eq_Cangular}
\begin{split}
\frac{d C_{\mathrm{angular}}}{d\lambda} = 0
\quad \Longrightarrow \quad
\left\{
\renewcommand{\arraystretch}{2.5}
\begin{array}{@{}l@{}}
\displaystyle
\frac{d}{d\lambda}
\left(
\sigma_{angular}\omega
\left(
r^2 \cos(\phi)u^\theta
- r^2 \cos(\theta)\sin(\theta)\sin(\phi)u^\phi
\right)
\right) = 0,
\\[0.6em]
\displaystyle
\frac{d}{d\lambda}
\left(
\sigma_{angular}\omega
\left(
r^2 \sin(\phi)u^\theta
+ r^2 \cos(\theta)\sin(\theta)\cos(\phi)u^\phi
\right)
\right) = 0,
\\[0.6em]
\displaystyle
\frac{d}{d\lambda}
\left(
\sigma_{angular}\omega
\left(
r^2 \sin^2(\theta)u^\phi
\right)
\right) = 0.
\end{array}
\right.
\end{split}
\end{equation}
Since $\sigma_{angular}$ is a constant, it can be pulled out of the derivative. We notice that the three conserved quantities in the above equation are our $L_x$, $L_y$ and $L_z$, if we include the $\omega$-term as well by defining $L_i \longrightarrow \omega L_i$ and call it $\tilde L_i$. Thus, the angular momentum $\tilde L_i$ is a conserved quantity. 
%%%%%%%%%%%%%%%%%%%
\subsection{Conserved quantities directly from $q$-desics}
In the following section, we determine the conserved quantities, energy and angular momentum, of the particle along $q$-desics using $q$-desics directly.
\subsubsection{Energy}
For the energy, we go to the first component of (\ref{eq_GeodSym})
\be
\frac{du^t}{d\lambda} + \langle \hat \Gamma^t_{\alpha \beta} \rangle u^\alpha u^\beta + u^t \frac{\dot \omega}{\omega } = 0 \longrightarrow \frac{dr}{d\lambda}\frac{du^t}{dr} + 2 \langle \hat \Gamma^t_{t r} \rangle u^t u^r + u^t\frac{1}{\omega } \frac{dr}{d \lambda}\frac{d \omega}{dr} = 0
\ee
Using $\Gamma^t_{t r}  = \langle \hat g' / g \rangle$, $u^r = dr/d\lambda$ and $\dot \omega = \frac{d \omega}{d\lambda} = \frac{dr}{d \lambda}\frac{d \omega}{dr} = u^r \frac{d \omega}{dr} $ and also dividing the equation by $u^r$, we obtain
\be
\frac{du^t}{dr} + \left\langle \frac{\hat g'}{\hat g} \right\rangle u^t + \frac{u^t}{\omega}\frac{d \omega}{dr } = 0
\ee
Again, using  the techniques discussed in Ref.~\cite{Koch:2025qzv}, we can write $\left\langle \frac{\hat g'}{\hat g} \right\rangle = \frac{1}{r} \left( \left\langle \frac{1}{\hat g} \right\rangle \left( 1-\Lambda r^2 \right)-1 \right) =A(r)$. The differential equation then becomes
\bea
&& \frac{d u^t}{d r} + u^t \frac{1}{r} \left( \left\langle \frac{1}{\hat g} \right\rangle \left( 1-\Lambda r^2 \right)-1 \right) + \frac{u^t}{\omega} \frac{d \omega}{dr} = 0
\\ \nonumber & \longrightarrow & \frac{d u^t}{d r} + u^t A(r) + \frac{u^t}{\omega} \frac{d \omega}{dr} = 0
\eea
One can simply solve this differential equation and obtain
\be
\ln{u^t}= \bar{c} - \ln{ \omega} - \int dr A(r) \longrightarrow u^ t= E \frac{1}{\omega} e^{- \int dr A(r)}
\ee
where we have defined $e^{\bar{c}}=E$. We immediately notice that this gives us the same equation for the energy as in eq.~(\ref{eq_energy_qilling_v1}),
\be\label{eq_EnFromQD}
E = u^t\omega e^{ \int dr A(r)}
\ee
Thus, the energy calculated here is the same as calculated using conserved quantities and $q$-illing vectors.
To see that this expression is indeed gauge invariant, it is instructive to start from an affine gauge, say with the parameter $\gamma$ where $\omega =1$.
In this gauge, $E\propto d t/d\gamma$.
If we now go to any other parameter $\lambda$, this is
$E \propto dt/d^\lambda \cdot d\lambda/d \gamma$. Now, we recall that $\omega$ represents the change of a new parameter with respect to the old $\omega =d\lambda/d \gamma $, which is exactly (\ref{eq_EnFromQD}).

\subsubsection{Angular Momentum}
For the angular momentum, we solve the $\theta$ and the $\phi$ components of the $q$-desics equation~\ref{eq_GeodSym}.
\be \label{eq_phi_ddot}
\ddot \phi + \frac{2 \dot r \dot \phi }{r} + 2 \cot ( \theta ) \dot \theta \dot \phi + \frac{\dot \omega}{\omega} \dot \phi = 0
\ee
\be \label{eq_theta_ddot}
\ddot \theta + \frac{2 \dot r \dot \theta }{r} - \cos \theta \sin \theta \dot \phi^2 + \frac{\dot \omega}{\omega} \dot \theta = 0
\ee
Multiplying the equation for $ \phi$  by $ r^2 \sin^2 (\theta) \omega$, we can write it as 
\be
\frac{d}{d \lambda} \left(r^2 \sin^2 (\theta) \dot \phi \omega \right) = 0
\ee
Hence $r^2 \sin^2 (\theta) \dot \phi  \omega$ is a conserved quantity. This quantity is just our $\omega L_z$ or the previously defined $\tilde \omega L_z$ and we see that the analysis from $q$-desics also results in the conservation of this quantity just as in the case with effective $q$-illing covectors. Now we would like to check the conservation of the other two angular momentum $L_x$ and $L_y$.

Taking the derivative of $L_x$ with respect to the affine parameter $\lambda$, we obtain
\bea
\frac{d}{d \lambda}L_x &=& \frac{d}{d \lambda} \left( -r^2 \sin ( \phi ) \dot \theta - r^2 \sin ( \theta ) \cos ( \theta ) \cos ( \phi ) \dot \phi  \right) \\ \nonumber
&=& - 2 r \dot r \sin ( \phi ) \dot \theta - r^2 \dot \phi \cos ( \phi ) \dot \theta -r^2 \sin ( \phi ) \ddot \theta - 2 r \dot r \sin ( \theta ) \cos ( \theta ) \cos ( \phi ) \dot \phi \\ \nonumber  & - &   r^2 \theta \cos^2 ( \theta ) \cos ( \phi ) \dot \phi + r^2 \dot \theta \sin^2 ( \theta ) \cos ( \phi ) \dot \phi + r^2 \sin ( \theta ) \cos ( \theta ) \dot \phi^2 \sin ( \phi ) - r^2 \sin ( \theta ) \cos ( \theta ) \cos ( \phi ) \ddot \phi 
\\ \nonumber
& = & \frac{\dot \omega }{\omega}  \left(  r^2 \dot \theta \sin ( \phi ) + r^2 \dot \phi \sin ( \theta ) \cos ( \theta ) \cos ( \phi ) \right) = - \frac{\dot \omega }{\omega}L_x %= \frac{d}{d \lambda} L_x
\eea
where in the last step we have used the equations~(\ref{eq_phi_ddot},\ref{eq_theta_ddot}). If we multiply both sides of the last equality with $\omega$, we can rewrite the last equality as 
\be
\omega\frac{d}{d \lambda} L_x + \dot \omega L_x = 0 \longrightarrow \frac{d}{d \lambda} \left( \omega L_x \right) = 0
\ee
We see that the quantity $\tilde L_x = \omega L_x$ is conserved just as in the case with q-killing vectors.

For $L_y$, one can do the same steps as for $L_x$ and see that the quantity $\tilde L_y = \omega L_y$ is conserved.
%%%%%%%%%%%%%%%%%%%%%

%%%%%%%%%%%%%%%%%
\section{Discussion} \label{disc}

%%%%%%%%%%%%%%%%%%%
\subsection{Is the proper time still an affine parameter?}
Our definition of proper time (\ref{eq_ProperTime}) involves only  expectation values 
of metric operators $\hat g_{\alpha \beta}$
and not their products or derivatives. 
The same is true for the conservation of the normalised velocity square (\ref{eq_U2conserved}).
Thus, according to this, we do not have to fear (or hope) that $q$-desic induced QG effects alter the relation between changes in proper time  and coordinate displacements e.g.
\be\label{eq_PropTime}
d\tau^2= dt^2 \langle \hat g_{00} \rangle+
dr^2
 \langle \hat g_{11} \rangle
 +d\theta^2 
  \langle \hat g_{22} \rangle
  +d\phi^2
  \langle \hat g_{33} \rangle.
\ee
However, QG aspects still affect time measurements indirectly. 
On the one hand they may change
the metric expectation values themselves.
On the other hand, and this is the novel aspect, $q$-desics will affect e.g. how we associate the conserved quantities (e.g. energy or angular momentum) to a time measurement in an experimental setup that involves free fall or orbital motion. \\
Even though (\ref{eq_PropTime}) is defined with the expectation value of the metric only, we cannot a-priory assume that it is also an affine parameter. For any parameter $\lambda$ describing a $q$-desic curve, there is in principle the $\dot \omega/\omega$ term on the right hand side. Only if the gauge fixing condition (choice of the evolution parameter) implies $\dot \omega=0$, we have an affine parameter.
Let's check this explicitly for the proper time choice. The associated gauge fixing condition is $
U^2=1$ (\ref{eq_U2conserved}).
Naturally, this implies
\be\label{eq_dU20}
\frac{d}{d\tau} U^2=0.
\ee
We can work this condition out by using the $q$-desic equation, while leaving the gauge parameter $\omega$ still undetermined.
We find that (\ref{eq_dU20}) implies
\be\label{eq_propGF}
\frac{\dot \omega}{\omega}= - U^\alpha U^\beta U_\nu \Delta^\nu_{\alpha \beta},
\ee
in analogy to the gauge fixing (\ref{eq_gaugefix2}). 
Thus, we can conclude that within the $q$-desic framework: {\it  ``proper time is not always an affine parameter''}.
%
%%%%%%%%%%%%%%%%%%%
\subsection{How does the $q$-desic equation read in proper- and in coordinate-time?}
With the discussion from the previous subsection, we are equipped to formulate the $q$-desic equations in an explicit gauge choice.
The most natural and frequent choices are coordinate time and proper time.
\begin{itemize}
    \item Proper time:\\
We can insert the condition (\ref{eq_propGF}) into the generic equation (\ref{eq_Covqdesic}) providing us the self-consistent $q$-desic equation in proper time gauge
\be\label{eq_Proptqdesic}
\frac{d U^\mu}{d\tau}=
-\Gamma^\mu_{\alpha \beta}(\langle \hat g \rangle) U^\alpha U^\beta
-\Delta^\nu_{\alpha \beta}U^\alpha U^\beta(\delta_\nu^\mu-U_\nu U^\mu),
\ee
where the QG correction is projected by $(\delta_\nu^\mu-U_\nu U^\mu)$ onto velocity components that are orthogonal to $U^\mu$.
\item Coordinate time:
Now we proceed with the formulation in coordinate time, which yields the velocities
\be
v^\mu= \frac{dx^\mu}{dt}
\ee
with the gauge condition
\be
v^0=1.
\ee
This gauge implies that the LHS of the first $q$-desic equation has to vanish $d v^0/dt=0$, which leads to
\be\label{eq_tGF}
\frac{\dot \omega}{\omega}=-\left\langle 
\hat \Gamma^0_{\alpha \beta}
\right\rangle v^\alpha v^\beta.
\ee
With this, the $q$-desic equations for the remaining spatial velocities in coordinate time read
\be\label{eq_qdesict}
\frac{d v^i}{dt}
=-\left(\left\langle 
\hat \Gamma^i_{\alpha \beta}
\right\rangle - v^i 
\left\langle 
\hat \Gamma^0_{\alpha \beta}
\right\rangle \right)v^\alpha v^\beta.
\ee
\end{itemize}
When we want to solve the $q$-desics in one of the two natural gauge choices (\ref{eq_Proptqdesic}, \ref{eq_qdesict}) it
is convenient to invoke the conserved quantities (\ref{eq_conservedq}), if $q$-illing vectors are available.
The definitions of the conserved quantities however, contain the gauge function $\omega$. Thus, we first have to derive this function by integrating the respective gauge conditions (\ref{eq_propGF}, \ref{eq_tGF}).

%%%%%%%%%%%%
\subsection{Gauge (in)-dependence}
\label{subsec_Gauge}

In systems with gauge invariance,
equations of motion can depend on the gauge parameter, while 
physical observables
should not depend on the gauge choice. The $q$-desics are no exception from this general scheme.
In the case of the $q$-desic equations, the gauge freedom is implemented in terms of the choice of the function $\omega$.
As it is to be expected, the dynamical equations (\ref{eq_qdesic}) do depend explicitly on $\omega$. 
What about the quantities
(\ref{eq_U2conserved}) and (\ref{eq_conservedq})?
The conserved quantity, (\ref{eq_conservedq})
takes the same value for any parameter $d\lambda$ and is thus by all means gauge invariant.
However, the normalised velocity vector in (\ref{eq_U2conserved}) is explicitly defined with respect to the proper time parameter $d\tau$, which is a particular gauge choice for $d\lambda$ and as we have learned from (\ref{eq_dU20}), $u^2$ is not a constant of motion for arbitrary velocities $u^\mu = d x^\mu /d\lambda$ with arbitrarily chosen $d\lambda$. It may be true
(e.g. in inertial frames), but it is only guaranteed for velocities $U^\mu=dx^\mu/d\tau$ defined with respect to the proper time of the test particle.
Therefore, we still can define a conserved mass and four momentum as in (\ref{eq_defp}), but this definition does involve the proper time and is thus not strictly speaking gauge invariant.
This may have some interesting observational consequences.

%%%%%%%%%%%%%%
\subsection{Classical limit of the orbital energy}

The q-energy (\ref{eq_EnFromQD}) involves an exponential of an integral over  a radial function which looks very different from the classical result (\ref{eq_conservedCl}) which basically just contains the metric function $g_{00}$.
So how is the classical limit recovered?

The answer to this question lies in the ordering of expectation values and operator products.
The function $A(r)$ in the integral within the exponential was defined as
\be
A(r)= \left\langle 
\frac{\hat g '}{\hat g}\right \rangle.
\ee
If we allow ourselves to approximate this expression as product of expectation values we find
\be
A(r)\approx   
\frac{\langle\hat g '\rangle}{\langle\hat g\rangle}=
\frac{d(\langle\hat g \rangle)/dr}{\langle\hat g\rangle}
=
\frac{d}{dr}
\ln (\langle \hat g_{..}\rangle).
\ee
Thus,
\be\label{eq_EnFromQD}
E = u^t\omega \,e^{ \int dr A(r)}
\approx 
u^t \omega \exp (\ln (\langle \hat g \rangle))=
u^t \omega \,\langle \hat g \rangle.
\ee
Note that naively we would have expected to get 
$\langle \hat g_{00} \rangle=\langle \hat n^2 \hat g\rangle $ and not just $\langle \hat g \rangle$ on the right and side.
However, this difference does not manifest itself in the classical limit since in the Schwarzschild solution, the differential $dt$ of the time-coordinate   is typically chosen such that $n^2=1$.

\subsection{Effective $q-$illing equation and quantum reference frames}
An intriguing conceptual parallel exists between the effective q-illing construction and quantum reference frame (QRF) approaches. In Eqs.~(\ref{eq_effective_qilling_ansatz},\ref{eq_Qilling}), the geometry entering through $\hat g_{\mu\nu}$ and $\hat\Gamma^{\alpha}_{\mu\nu}$ remains quantum, whereas the effective q-illing covector $\zeta_\mu$ is a c-number field on the underlying classical manifold. QRF formulations exhibit a related structure: reference systems are treated as quantum systems, while choosing a particular perspective can lead to relational coordinate descriptions associated with ordinary c-number labels or symmetry parameters \cite{Giacomini2019QRF,Vanrietvelde2020Perspective,Kabel2023Boundary,deLaHamette2023IndefiniteMetric,HausmannRenner2026ThirdParticle}. This suggests that $\zeta_\mu$ may admit an interpretation as an effective relational symmetry descriptor determined by quantum-geometric data. We stress, however, that this is presently only a structural analogy; establishing a genuine QRF interpretation would require deriving Eq.~(\ref{eq_Qilling}) from an explicit relational conditioning or reference-frame reduction rather than from the ansatz $\hat\xi_\mu\simeq\zeta_\mu\mathbf{1}$. Whether $\zeta_\mu$ actually exhibits such a nature and whether there exists such a deep and genuine relation to QRF remains an open question. We will address this more precisely in a future work.

\subsection{The operator $q$-illing equation and quantum symmetry sectors}

Equation~(\ref{eq_Killing2}) may admit an interpretation beyond that of a direct
operator-valued generalisation of the classical Killing equation. It is
useful to introduce the $q$-illing operator
\begin{equation}
    \hat{\mathcal K}_{\mu\nu}[\xi]
    \equiv
    \hat{\nabla}_{\mu}\hat{\xi}_{\nu}
    +
    \hat{\nabla}_{\nu}\hat{\xi}_{\mu}
    =
    \left(\hat{\mathcal L}_{\xi}\hat g\right)_{\mu\nu}.
\end{equation}
In the present work, the $q$-illing condition
$\hat{\mathcal K}_{\mu\nu}[\xi]=0$ is understood as a symmetry condition on the operator-valued geometry. This should be distinguished from the constraints of canonical gravity, such as the Hamiltonian constraint, which arise as first-class constraints and, after quantisation, are imposed on physical states in the Dirac--Wheeler--DeWitt sense
\cite{DeWitt1967,Kuchar1993Canonical,Thiemann2007}.

Nevertheless, Eq.~(\ref{eq_Killing2}) may suggest a related, and potentially useful, state-dependent notion of quantum spacetime symmetry. For a given candidate symmetry field $\xi^\mu$, one may consider the condition $ \hat{\mathcal K}_{\mu\nu}[\xi]\,|\Psi\rangle=0$,
which would select a sector of quantum-geometrical states possessing the corresponding exact $q$-illing symmetry. Formally, one may define $ \mathcal H_{\xi} = \left\{ |\Psi\rangle\in\mathcal H_g\,\big|\, \hat{\mathcal K}_{\mu\nu}[\xi]|\Psi\rangle=0\right\}$.
Equation $ \hat{\mathcal K}_{\mu\nu}[\xi]\,|\Psi\rangle=0$ has a structure reminiscent
of quantum constraint equations, although at present it should be viewed
as a symmetry-sector condition rather than as an additional fundamental
constraint of the theory. In particular, the classical Killing equation
does not itself constitute a Dirac constraint. A related connection
between Killing symmetries and canonical initial data is provided by the
Killing-initial-data construction \cite{BeigChrusciel1997KID}.

This viewpoint also distinguishes different levels at which a quantum
geometry may possess a symmetry. One may require the $q$-illing condition
as an operator identity,
\begin{equation}
    \hat{\mathcal K}_{\mu\nu}[\xi]=0,
\end{equation}
as a condition on a particular state,
\begin{equation}
    \hat{\mathcal K}_{\mu\nu}[\xi]|\Psi\rangle=0,
\end{equation}
or only at the level of expectation values,
\begin{equation} \label{eq_qillingConstraint_VEV}
    \langle\Psi|
    \hat{\mathcal K}_{\mu\nu}[\xi]
    |\Psi\rangle=0 .
\end{equation}
The latter condition need not imply an exact quantum symmetry, since fluctuations such as
$\langle
\hat{\mathcal K}_{\mu\nu}^{\dagger}
\hat{\mathcal K}_{\mu\nu}
\rangle$ may remain non-vanishing. This suggests a possible distinction between exact operator symmetries, state-dependent quantum symmetries, and symmetries which emerge only after taking expectation values. The effective $q-$illing covector equation in our work, namely eq.~\ref{eq_Qilling}, emerges from eq.~\ref{eq_qillingConstraint_VEV} by using the ansatz eq.~\ref{eq_effective_qilling_ansatz}. However, more detailed studies of this interpretation of the $q-$illing operator and its implications for the quantum-geometrical states are required, which is beyond the scope of this work. We will address this more clearly in a future work.

%%%%%%%%%%%%%%%

%%%%%%%%%%%%%%%%%%%%%%%%%%%%%
\subsection{Null-$q$-desics?}

The above discussion applies to test particles with finite proper time, which cannot be defined along a null curve. Consequently, when testing whether $(d/d\lambda)u^2$ vanishes, the gauge choice (\ref{eq_gaugefix2}) is unavailable because the $\dot\omega/\omega$ term is multiplied by $u^2=0$. This leaves several possible scenarios:
\begin{itemize}
\item The motion of structureless massless test particles is not by $q$-desics, but instead by geodesics or some other formulation or effect~\cite{Parikh:2020kfh,Freidel:2026hed}. 
\item 
Massless structureless test particles follow $q$-desics, allowing the following sub-scenarios:
\begin{itemize}
    \item The QG corrections $\Delta^\mu_{\alpha \beta}$ always have a structure such that
    \be\label{eq_QGcorrm2}
    \Delta^\mu_{\alpha \beta} U_\mu U^\alpha U^\beta=0,
    \ee
    for any null vector $U^\mu$. This would happen for example if
    $\Delta^\mu_{\alpha \beta} \propto A^\mu \langle g_{\alpha \beta}\rangle$,
    where $A^\mu$ is an arbitrary 4-vector.
 %   The condition (\ref{eq_QGcorrm2}) would reduce the 64 possible functions in $\Delta^\mu_{\alpha \beta}$ to 24 \textcolor{blue}{(??? basically a anti-symmetric 2x2-tensor ($10-4=6$) times a 4-vector: $6\cdot 4=24$ ???add discussion in appendix ???)} possible independent functions.
    In this case, $q$-desics would simply preserve null curves, independent of the parameter choice~$\lambda$.
    \item 
    The QG corrections do not give (\ref{eq_QGcorrm2}). This would imply that they can take curves that are initially null into curves that are either space-like or time-like. This type of ``mass generation'' is frequently anticipated in discussions on possible effects of QG~\cite{Saravani:2015rva,Belenchia:2014fda,Amelino-Camelia:2011lvm}.
\end{itemize}
\item Treating particles as structureless is only approximate, since neglecting spin already omits $\mathcal O(\hbar)$ effects. Thus, including QG-induced $\mathcal O(\hbar)$ corrections while ignoring spin may be inconsistent with $u^2=0$. A consistent treatment should include both on equal footing~\cite{Gambini:1998it,Armaza:2016vfz,Asenjo:2016hnc}, for instance through massless spinning particles and their $q$-desic generalisations or through an eikonal/WKB approximation of the corresponding field theory \cite{Mieling:2026wsv}.
\end{itemize}
We postpone the investigation of massless-$q$-desics in the above scenario. This question deserves a focused discussion, dedicated explicitly to this problem.

%%%%%%%%%%%%%%%%%%%%%%%%%%%%%%%%%%
\section{Conclusion}
We have shown that the classical trinity between:
\begin{center}
    {\it Killing symmetries of spacetime} $\leftrightarrow$
{\it geodesic motion} $\leftrightarrow$ {\it conserved charges}
\end{center} can be
formulated for the $q$-desic framework. 
In particular, we showed that the effective $q$-illing covectors defined by (\ref{eq_Qilling}) allows to define a corresponding conserved charge (\ref{eq_conservedq}).
Thus, we found that:
\begin{center}
   {\it $q$-illing symmetries of spacetime} $\leftrightarrow$
{\it $q$-desic motion} $\leftrightarrow$ {\it conserved charges}, 
\end{center}
can be consistently implemented.
We then gave concrete examples for these quantities for the case of a time-independent QG background with spherical symmetry, where we derived the conserved energy and angular momentum for the $q$-desic approach.

We further gave a precise definition of the $q$-desic motion when described in terms of proper time $d\tau$ and in terms of coordinate time $dt$. In this context we showed that the definition of a conserved mass explicitly involves the velocity defined in terms of displacements with respect to this proper time.

First and foremost, our findings underline the self-consistency of the $q$-desics framework.
In addition, they will allow us on the one hand to solve $q$-desics in symmetric backgrounds much easier. On the other hand, they will also make is possible to interpret and understand such solutions and their physical meaning much more precisely.

%\textcolor{red}{TODO}

%%%%%%%%%%%%%%%%%%%%%%%%%%%%%%%%%%
\begin{acknowledgments}
We thank L. Freidel, L. Heisenberg, J. Isidro, V. Kabel, C. Pfeiffer, and R. Renner for valuable discussion in the process of preparing this work.
A.~R.\ would like to express his gratitude to Silesian University in
Opava for financial support, and acknowledges the Departamento de
F\'isica of Universidad del B\'io-B\'io.
The creation of this article was supported by the grant program
Vouchers for Universities in the Moravian-Silesian Region
(registration number CZ.10.03.01/00/23\_042/00003901119).
This article is based upon work from COST Action FuSe, CA24101,
supported by COST (European Cooperation in Science and Technology).
\end{acknowledgments}

{\bf{Conflicts of Interest: }}The authors declare no conflict of interest.\\

{\bf{Data Availability Statement: }}There is no data available for this research.\\

%{\bf{Funding: }}This research received no external funding.

%%%%%%%%%%%%%%%%%%%%%%%%%%%%%%%%%%
%\newpage
\appendix

\section{Potentially divergent $1/\hat g$ operator}\label{sec_divergencies}

The equations for the spherically symmetric system (\ref{eq_GeodSym}) contain inverse operators $ 1/\hat g$. This can lead to divergencies when we try to evaluate expectation values like $\langle 1/\hat g \rangle$. At this point, we started exploring three strategies that allow to cope with this potential problem:
\begin{itemize}
    \item {\bf Weak field expansion}\\
    In a regime of weak gravitational effects one can argue that $\hat g$ is close to the unit operator and formally expand in this vicinity
    \be\label{eq_ginvEV}
%\langle \hat g^{-1} \rangle \approx 3 - 3 \langle \hat g \rangle - \langle \hat g^{2} \rangle
\langle \hat g^{-1} \rangle = \sum_{n=1}^\infty \sum_{l=0}^{n-1}  \left( \begin{array}{c}
         n-1  \\
         l 
    \end{array}\right)
      (-1)^{l} \langle \hat g^l \rangle.
\ee
\item {\bf Regularisation}\\
In a Schr\"odinger picture, where the spacetime wave functions $\psi(g,r)$ are available, one can regularise the divergent integral 
\be
\langle 1/\hat g \rangle =
\int_{-\infty}^{+\infty} dg \;\psi^* \frac{1}{g}\psi
\ee
by standard techniques like a complex $i \epsilon$ shift of the pole and subsequently try to isolate and remove the divergent  from the physical part. 
\item {\bf Minimal multiplication}\\
The $q$-desics itself arises from the subtle order in which one performs operator products and the subsequent evaluation of expectation values. Thus, it is natural to use the same tool to circumvent the appearance of the divergent $1/\hat g$ operator in the first place. This can be achieved by multiplying the entire relation (\ref{eq_GeodSym}) with the minimal necessary power $\hat g^1$, before taking the EV. Then, after canceling operators with their inverse, we perform the Weyl ordering and take the EV. Finally we divide the relation by $\langle \hat g \rangle$. This results in
\be\label{eq_GeodSymReg}
\left(
\begin{array}{c}
   \dot  u^t  \\
   \dot  u^r\\
   \dot  u^\theta\\
   \dot  u^\phi
\end{array}
\right)
=
\left(
\begin{array}{c}
-\langle \hat g'\rangle/\langle \hat g\rangle u^r u^t-\frac{\dot \omega}{\omega}u^t\\
(u^r)^2 \langle\hat g'\rangle/\langle 2 \hat g\rangle+\left\langle\hat g^2\left(r (u^\theta)^2+r \sin^2\theta (u^\phi)^2-  \hat n^2 \hat g' (u^t)^2/2 \right)\right\rangle/\langle \hat g\rangle-\frac{\dot \omega}{\omega}u^r\\
- 2 u^r u^\theta /r + \cos \theta \sin \theta (u^\phi)^2-\frac{\dot \omega}{\omega}u^\theta\\
- 2(u^r u^\phi+ r \cot \theta u^\theta u^\phi)/r-\frac{\dot \omega}{\omega}u^\phi
\end{array}
\right)
\ee
which has a regularised connection defined by
\be
\langle \hat \Gamma^\mu_{\alpha \beta} \rangle_{MM}\equiv \frac{
\langle \hat  \Gamma^\mu_{\alpha \beta} \hat g \rangle}{\langle \hat g \rangle}.
\ee
\end{itemize}
The appearance of potentially divergent quantities is nothing distinctive of $q$-desics.
We plan to explore the above ideas and their relation to existing renormalisation schemes in our future work, keeping in mind that a true physical prediction should not depend on the method we used for it's calculation.

%%%%%%%%%%%%%%%%%%%%
%%%%%%%%%%%%%%%%
\section{Formalism: Manifold, Configuration Space and the Hilbert Space}\label{sec_app_formalism}

In our work, we take a 4-dimensional, smooth and classical manifold $\mathcal{M}$ with spacetime points $P \in \mathcal{M}$ and local basis $ \{ \partial_{\mu} \} $. On a coordinate patch $U\subset \mathcal{M}$, we introduce local coordinates
\be
    x^\mu:U\to\mathbb R,\qquad \mu=0,1,2,3.
\ee
Thus, a spacetime point is denoted by $P \in \mathcal{M}$, while $x^\mu(P)$ are its coordinate values in a chosen chart.

%We further assume that on the manifold $ \{ \partial_{\mu} \} $ the coordinates $x^{\mu}$ are defined and exist.The spacetime points $P$ are labeled by $x^{\mu}$.
%, $x^{\mu} \in \mathcal{M}$.

%The configuration space, which is the space in which the metrics are defined, is defined on this smooth manifold $\mathcal{M}$, $\mathcal{S}_{metric}= \{ all \; g_{\mu \nu} \; defined \; on \; \mathcal{M}  \} $.
Before symmetry reduction, one may think of the classical gravitational configuration space schematically as the set of admissible metric fields on $\mathcal{M}$,
\be
\mathcal S_{metric}=\{g_{\mu\nu}\ \text{admissible on } \mathcal{M} \}.
\ee
This is not the spacetime manifold itself, but the space of possible metric configurations on that manifold.

In our framework, we do not work with the full configuration space of arbitrary metric fields. Instead, we restrict to a reduced sector in which the geometry is parametrised by the reduced metric function $g(r)$. We do this by imposing static spherical symmetry. Our configuration variable then is the metric function $g$ considered as the canonical variable. The canonical variable $g$ together with its conjugate momentum $p$ form the phase space. Due to symmetry reduction, we do not work with the full configuration space. Our configuration space is the reduced configuration space in which all $g(r)$ are defined on $\mathcal{M}$,
\be
\mathcal S_{reduced}=\{ g(r) \ \text{admissible on } \mathcal{M} \}.
\ee
and this implies $\mathcal{S}_{reduced} \subset \mathcal{S}_{metric}$. Thus, so far we have
\be 
    P \in \mathcal{M} \quad \text{and } \quad g(r) \in \mathcal{S}_{reduced} \subset \mathcal{S}_{metric} 
\ee

We define our formalism before quantisation on a smooth manifold with metric function $g(r)$ as the configuration variable defined on the manifold $\{ g, \mathcal{M} \}$. After that, we promote the canonical pair to operators, $g \longrightarrow \hat g$ and $p \longrightarrow \hat p$ while the manifold is still classical and smooth.

For the quantisation, we define the Hilbert space as a space of square-integrable wave functions 
$\Psi(g)$ over the configuration variable $g$. The quantum states of the reduced gravitational sector are taken to belong to this space. The Hilbert space of the reduced gravitational sector is then defined as 
\be
\mathcal{H}_{g} = L^2(\mathcal{S}_{reduced}, d\mu(g)) = L^2(\mathcal{S}_{reduced},\mathcal{J}(g) dg)
\ee
equipped with the inner product
\be
\langle \Psi | \Phi \rangle =   \int_{\mathcal{S}_{reduced}} \Psi^*(g)\,\Phi(g)\,d\mu(g) 
\ee
This reduced Hilbert space is separable and thus admits 
a countable orthonormal basis $\{\phi_n(g)\}$ satisfying 
\be
\langle \phi_m | \phi_n \rangle =  \delta_{mn} 
\quad \text{and} \quad
\sum_n |\phi_n\rangle\langle \phi_n| = \mathbb{I}
\ee

In addition, we assume there exists a continuous set of generalised eigenstates $\{|g\rangle\}$ of the operator $\hat g$ in this reduced Hilbert space satisfying, $\hat g\,|g\rangle = g\,|g\rangle$, $\langle g|g'\rangle = \mathcal{J}(g)^{-1}\delta(g-g')$ and the completeness relation $\int_{\mathcal{S}_{reduced}} |g\rangle\langle g|\,d\mu(g)=\mathbb{I}$. In this representation, any state $|\psi\rangle$ admits the expansion $|\psi\rangle = \int_{\mathcal{S}_{reduced}} \psi(g)\,|g\rangle\,d\mu(g)$ with $\psi(g)=\langle g|\psi\rangle$. Formally, they provide a generalised spectral representation of states in the $g$-representation.\\

In constrained Hamiltonian systems, not all states are physical due to  the presence of the constraints. Therefore, we have to distinguish between kinematical Hilbert space $\mathcal{H}_{kinematical}$ and physical Hilbert space $\mathcal{H}_{physical}$ where the former is the Hilbert space in which the constraints are not considered and the latter is the Hilbert space in which the constraints are satisfied, $\mathcal{H}_{physical} \subset \mathcal{H}_{kinematical}$. 

In cases where the constraints are of first class, one imposes the condition that the physical Hilbert space is defined by acting with the constraints on the wave functional such that they vanish,
\begin{equation}
    \hat \chi_a | \Psi(g)\rangle = 0
\end{equation}
where $\hat \chi_a $ are the first class constraints promoted to operators after quantisation. Therefore, the Hilbert space with the presence of the first class constraints is written as $\mathcal{H}_{physical}= \{ \Psi \in \mathcal{H}_{kin} \; \big| \; \hat \chi_a | \Psi(g)\rangle = 0  \}$.  In our work however, we have second class constraints only and this condition does not work when having second class constraints only. When there are second class constraints, one first goes from Poisson brackets to Dirac brackets. The second class constraints then become strongly zero. The quantisation is then performed on the constraint surface.
Since in the present reduced treatment the second-class constraints are first taken into account at the classical level through Dirac brackets and are then imposed strongly on the reduced phase space, we do not distinguish further between kinematical and physical Hilbert space. Accordingly, in this work we simply write

%Hence, in our work and due to the presence of second class constraints, the kinematical Hilbert space is already the physical Hilbert space. 

\be
    \mathcal{H}_{physical} \equiv \mathcal{H}_{kinematical} \equiv \mathcal{H}_{g} 
\ee

In summary, in the present framework, the differentiable spacetime manifold $\mathcal{M}$ is kept classical and smooth. Spacetime points $P\in\mathcal M$ are assigned coordinate values $x^\mu(P)$ in a chosen chart. The quantum aspect enters through the promotion of the reduced canonical pair $g\to\hat g$ and $p\to\hat p$ acting on a reduced gravitational Hilbert space $\mathcal{H}_{reduced}$ while keeping the manifold smooth and classical $\{ g, \mathcal{M} \} \longrightarrow \{ \hat g , \mathcal{M} \}$ and $\{ g, p\} \longrightarrow \{ \hat g , \hat p \}$. The Greek indices refer to the classical manifold while the hat refers to the object being promoted to an operator acting on the Hilbert space $\mathcal{H}_{g}$. The generalised eigenstates $| g \rangle $ provide a generalised spectral representation of states in the $g-$representation.

\section{Derivation of the effective $q$-illing vectors of a static spherically symmetric spacetime}
\label{appendix_qillingvectors}

Equation (\ref{eq_Qilling}) gives us ten differential equations. In order to find $\zeta_{\mu}$, we need to solve these equations. For static spherically symmetric spacetime the only non-vanishing Christoffel symbols are $\hat \Gamma^{0}_{10}$, $\hat \Gamma^{0}_{01}$, $\hat \Gamma^{1}_{00}$, $\hat \Gamma^{1}_{11}$, $\hat \Gamma^{1}_{22}$, $\hat \Gamma^{1}_{33}$, $\hat \Gamma^{3}_{23}$, $\hat \Gamma^{3}_{32}$, $\hat \Gamma^{2}_{12}$, $\hat \Gamma^{2}_{21}$, $\hat \Gamma^{2}_{33}$, $\hat \Gamma^{3}_{13}$, $\hat \Gamma^{3}_{31}$. Hence, the ten differential equations are given as
\begin{equation}
\label{eq_killing_eq1}
\partial_t \zeta_t - \langle \hat \Gamma^{r}{tt} \rangle \zeta r = \partial_t \zeta_t - \frac{1}{2r^2\Gamma^2} \langle \hat g \hat p^2 \hat g' \rangle \zeta_r = \partial_t \zeta_t - \frac{1}{2r^3\Gamma^2} \langle \hat g \hat p^2 \rangle \zeta_r + \frac{\Lambda}{2r \Gamma^2} \langle \hat g \hat p^2 \rangle \zeta_r + \frac{1}{2r^3 \Gamma^2} \langle \hat g \hat p^2 \hat g \rangle \zeta_r= 0
\end{equation}
\begin{equation}
\label{eq_killing_eq2}
\partial_r \zeta_r - \langle \hat \Gamma^{r}{rr} \rangle \zeta_r = \partial_r \zeta_r + \frac{1}{2} \left\langle \frac{\hat g'}{\hat g} \right\rangle \zeta_r = \partial_r \zeta_r + \frac{1}{2r} \left\langle \frac{1}{\hat g} \right\rangle \zeta_r - \frac{r\Lambda}{2} \left\langle \frac{1}{\hat g} \right\rangle \zeta_r- \frac{1}{2r} \zeta_r = 0
\end{equation}
\begin{equation}
\label{eq_killing_eq3}
\partial\theta \zeta_\theta - \langle \hat \Gamma^{r}{\theta\theta}\rangle \zeta_r = \partial\theta \zeta_\theta + r \langle \hat g \rangle \zeta_r = 0
\end{equation}
\begin{equation}
\label{eq_killing_eq4}
\partial_\phi \zeta_\phi - \langle \hat \Gamma^{r}{\phi\phi} \rangle \zeta_r - \langle \hat \Gamma^{\theta}{\phi\phi} \rangle \zeta_\theta = \partial_\phi \zeta_\phi + r \sin^2(\theta) \langle \hat g \rangle \zeta_r + \cos(\theta) \sin(\theta) \zeta_\theta = 0
\end{equation}
\begin{equation}
\label{eq_killing_eq5}
\partial_r \zeta_t + \partial_t \zeta_r - 2 \langle \hat \Gamma^{t}{rt} \rangle \zeta_t = \partial_r \zeta_t + \partial_t \zeta_r - \left\langle \frac{\hat g'}{\hat g} \right\rangle \zeta_t = \partial_r \zeta_t + \partial_t \zeta_r - \frac{1}{r} \left\langle \frac{1}{\hat g} \right\rangle \zeta_t + r\Lambda\left\langle \frac{1}{\hat g} \right\rangle \zeta_t + \frac{\zeta_t}{r}= 0
\end{equation}
\begin{equation}
\label{eq_killing_eq6}
\partial_r \zeta\theta + \partial_\theta \zeta_r - 2 \langle \hat \Gamma^{\theta}{r\theta} \rangle \zeta\theta = \partial_r \zeta_\theta + \partial_\theta \zeta_r - \frac{2}{r} \zeta_\theta = 0
\end{equation}
\begin{equation}
\label{eq_killing_eq7}
\partial_r \zeta_\phi + \partial_\phi \zeta_r - 2 \langle \hat \Gamma^{\phi}{r\phi} \rangle \zeta\phi = \partial_r \zeta_\phi + \partial_\phi \zeta_r - \frac{2}{r} \zeta_\phi = 0
\end{equation}
\begin{equation}
\label{eq_killing_eq8}
\partial_\theta \zeta_\phi + \partial_\phi \zeta_\theta - 2 \langle \hat \Gamma^{\phi}{\phi\theta} \rangle \zeta\phi = \partial_\theta \zeta_\phi + \partial_\phi \zeta_\theta - 2 \cot (\theta) \zeta_\phi = 0
\end{equation}
\begin{equation}
\label{eq_killing_eq9}
\partial_t \zeta_\phi + \partial_\phi \zeta_t = 0
\end{equation}
\begin{equation}
\label{eq_killing_eq10}
\partial_t \zeta_\theta + \partial_\theta \zeta_t = 0
\end{equation}
where we have used the equation of motion for $\hat g'$, i.e. $\hat g' = \frac{1}{r} \left( 1 - r^ 2\Lambda - \hat g \right)$, and we have also used $\hat n = \frac{\hat p }{r\Gamma}$.\

We allow for any coordinate dependency for the $\zeta_{\mu} \equiv \zeta_{\mu} (t,r,\theta \phi)$. Defining $A(r) = \frac{1}{r} \left( \langle \hat g^{-1} \rangle \left( 1-r^2\Lambda \right) -1 \right)$, we solve the (\ref{eq_killing_eq2}) as
\be
\partial_r \zeta_r +\frac{1}{2} A(r) \zeta_r = 0 \longrightarrow \zeta_r (t,r,\theta, \phi) = F(t,\theta, \phi ) R(r) = F(r,\theta, \phi) e^{-\frac{1}{2} \int dr A(r)}
\ee
Now we take the $r-$partial derivative of (\ref{eq_killing_eq3}) and $\theta-$partial derivative of (\ref{eq_killing_eq6}). Equating them, using $\partial_r \zeta_r = - \frac{1}{2}A(r) \zeta_r$ and using $\zeta_r (t,r,\theta, \phi) = F(t,\theta, \phi ) R(r)$, we obtain
\be
%-r \partial_r \langle \hat g \rangle \zeta_r - \langle \hat g \rangle \zeta_r + \frac{r}{2} \langle \hat g \rangle A(r) \zeta_r + \partial_{\theta}^2 \zeta_r + 2 \langle \hat g \rangl \zeta_r = 0
\partial_{\theta}^ 2 F(r,\theta, \phi ) + N(r) F(t,\theta , \phi) = 0 \quad \text{with} \quad N(r)= - r \partial_r \langle \hat g \rangle + \frac{r}{2} \langle \hat g \rangle A(r)
\ee
The function $N(r)$ does not vanish in general. Thus, $F(t,\theta,\phi)$ must be zero. We find

\be
\zeta_r (t,r,\theta, \phi) = 0
\ee
%This is as expected due to symmetry reasons and also in agreement with the radial killing vector in general relativity.
Now we set $\zeta_r$ to zero in the differential equations above. From (\ref{eq_killing_eq1}) and (\ref{eq_killing_eq3}) we infer that $\zeta_t$ is time independent and $\zeta_\theta$ is $\theta$ independent,
\be
\zeta_t(t,r,\theta, \phi) \rightarrow \zeta_t(r,\theta, \phi) \quad \text{and} \quad \zeta_\theta(t,r,\theta, \phi) \rightarrow \zeta_\theta(t,r, \phi)
\ee
Using (\ref{eq_killing_eq5}, \ref{eq_killing_eq6} and \ref{eq_killing_eq7}) we find
\be
\partial_r \zeta_t - A(r) \zeta_t = 0 \longrightarrow \zeta_t (r, \theta, \phi) = c_0 (\theta, \phi ) e^{\int dr A(r)}
\ee
\be
\partial_r \zeta_\theta - \frac{2}{r} \zeta_\theta = 0 \longrightarrow \zeta_\theta (t, r, \phi) = r^ 2 q_\theta (t, \phi)
\ee
\be
\partial_r \zeta_\phi - \frac{2}{r} \zeta_\phi = 0 \longrightarrow \zeta_\phi (t, r,\theta, \phi) = r^ 2 q_\phi (t, \theta, \phi)
\ee
Inserting $\zeta_t$ and $\zeta_\theta$ in (\ref{eq_killing_eq10}) and dividing by $r^2$ we get
\be
\partial_t q_\theta (t, \phi) + \frac{ e^{\int dr A(r)}}{r^2} \partial_{\theta} c_0 (\theta, \phi) = 0
\ee
The term $\partial_t q_\theta (t, \phi) $ is independent of $r$, and the term $\frac{ e^{\int dr A(r)}}{r^2} \partial_{\theta} c_0 (\theta, \phi)$ is $r-$dependent. Thus, each term must vanish separately. Therefore
\be
\partial_t q_\theta (t, \phi) = 0 \longrightarrow q_\theta (t, \phi) \equiv q_\theta (\phi) \quad \text{and} \quad \partial_{\theta} c_0 (\theta, \phi) = 0 \longrightarrow c_0 (\theta, \phi)\equiv c_0 (\phi)
\ee
Now we insert $\zeta_\phi$ and $\zeta_t$ in (\ref{eq_killing_eq9}) and obtain
\be
\partial_t q_\phi(t,\theta,\phi) + \frac{ e^{\int dr A(r)}}{r^2} \partial_{\phi} c_0 ( \phi)=0
\ee
Again, both terms must vanish separately and we get
\be
\partial_t q_\phi (t, \theta, \phi) = 0 \longrightarrow q_\phi (t,\theta, \phi) \equiv q_\theta (\theta,\phi) \quad \text{and} \quad \partial_{\phi} c_0 ( \phi) = 0 \longrightarrow c_0 ( \phi)\equiv c_0 = \text{constant}
\ee
In order to find $q_\theta$ and $q_\phi$ we use (\ref{eq_killing_eq8}) and (\ref{eq_killing_eq4}). Integrating (\ref{eq_killing_eq4}) over $\phi$ we get
\be
q_\phi(\theta, \phi) - h(\theta) + \cos (\theta) \sin (\theta) Q(\phi)= 0 \longrightarrow q_\phi(\theta, \phi) = h(\theta) - \cos (\theta) \sin (\theta) Q(\phi)
\ee
where $Q(\phi) = \int d\phi q_\theta(\phi)$ and $h(\theta)$ is the integration constant coming from integrating $\partial_{\phi} q_\phi(\theta,\phi)$ over $d\phi$.\
Using $q_\phi(\theta, \phi)$ and inserting it in (\ref{eq_killing_eq8}) we obtain
\be
\partial_{\theta} h (\theta) - 2 \cot (\theta) h(\theta) = - \partial^2_{\phi} Q(\phi) - Q (\phi)
\ee
The left hand side does not depend on $\phi$ and the right hand side does not depend on $\theta$. Hence, each side of the equation must be a constant.
\be
\partial^2_{\phi} Q(\phi) + Q (\phi) = \Tilde{C} \longrightarrow \int d\phi q_\theta (\phi) = Q(\phi) = c_1 \sin (\phi) - c_2 \cos (\phi) + \Tilde{C} \longrightarrow q_\theta(\phi) = c_1 \cos (\phi) + c_2 \sin (\phi)
\ee
\bea
    \partial_{\theta} h (\theta) - 2 \cot (\theta) h(\theta) = - \Tilde{C} & \longrightarrow & h(\theta) = \Tilde{C} \cos (\theta) \sin (\theta) + c_3 \sin^2 (\theta) \\ \nonumber & \longrightarrow & q_\phi (\theta , \phi)= c_3 \sin^2(\theta) - c_1 \cos ( \theta) \sin ( \theta) \sin ( \phi) + c_2 \cos ( \theta ) \sin ( \theta ) \cos ( \phi)
\eea
The solutions of $\zeta_t$, $\zeta_r$, $\zeta_\theta$ and $\zeta_\phi$ are then given by
\be
%\label{sol_zeta0_v1}
\zeta_t = c_0 e^{\int dr A(r)} \quad \text{with}\quad A(r) = \frac{1}{r} \left( \langle \hat g^{-1} \rangle \left( 1-r^2\Lambda \right) -1 \right)
\ee

\be
%\label{sol_zeta1_v1}
\zeta_r =0
\ee

\be
%\label{sol_zeta2_v1}
\zeta_\theta =r^2 \left( c_1 \cos (\phi) + c_2 \sin( \phi) \right)
\ee

\be
%\label{sol_zeta3_v1}
\zeta_\phi =c_3 r^2 \sin^2 (\theta) - r^2 \cos(\theta) \sin( \theta) \left( c_1 \sin (\phi) - c_2 \cos( \phi) \right)
\ee

\section{Inverse Operators in Weak Field Expansion}

We may need to have expressions for the  two inverse metric operators such as 
\be
    \left\langle \frac{1}{\hat n^2 \hat g^2} \right\rangle \quad \text{and} \quad \left\langle \frac{1}{\hat n^2 \hat g} \right\rangle
\ee
Let us define 
    \be
    \label{eq:inverse_lnO}
    \ln(\hat O)= - \sum_{n=1}^\infty\frac{\hat P^n}{n}
    \ee
    with $\hat P =(1-(\hat O))$. Then it follows
    \bea
    \left \langle\frac{1}{(\hat O)} \right \rangle = \left \langle\frac{\tilde \delta \ln ((\hat O))}{\tilde \delta (\hat O)} \right \rangle  =
    \sum_{n=1}^\infty
     \left \langle
     \hat P^{n-1}
     \right \rangle = 3-3\langle (\hat O)\rangle+\langle (\hat O)^2\rangle +...
     \label{eq_Oinv1} 
     \eea 
where we expanded to next to leading order of the small operator $\hat P$. Now defining for example $\hat O \in \{ \hat n^2 \hat g , \hat  n^2 \hat g^2 \}$, we can compute both of the desired inverse operators in series expansion just as we did before for the inverse of metric operator $\hat g$.

\bibliography{Ref_used.bib}

@article{Kuchar:1994zk,
    author = "Kuchar, Karel V.",
    title = "{Geometrodynamics of Schwarzschild black holes}",
    eprint = "gr-qc/9403003",
    archivePrefix = "arXiv",
    reportNumber = "UU-REL-94-3-1",
    doi = "10.1103/PhysRevD.50.3961",
    journal = "Phys. Rev. D",
    volume = "50",
    pages = "3961--3981",
    year = "1994"
}

@article{Mieling:2026wsv,
    author = "Mieling, Thomas B.",
    title = "{Internal dynamics and guided motion in general relativistic quantum interferometry}",
    eprint = "2602.21001",
    archivePrefix = "arXiv",
    primaryClass = "gr-qc",
    month = "2",
    year = "2026"
}

@article{Balsells:2025iac,
    author = "Balsells, Joseph and Bojowald, Martin",
    title = "{Quantum proper time: A Finsler space from entropy and purity}",
    eprint = "2503.06667",
    archivePrefix = "arXiv",
    primaryClass = "quant-ph",
    doi = "10.1103/2pn6-h5j4",
    journal = "Phys. Rev. D",
    volume = "113",
    number = "12",
    pages = "124060",
    year = "2026"
}

@article{Balsells:2024lpl,
    author = "Balsells, Joseph and Bojowald, Martin",
    title = "{Geometry and proper time of a relativistic quantum clock}",
    eprint = "2410.08156",
    archivePrefix = "arXiv",
    primaryClass = "gr-qc",
    doi = "10.1103/PhysRevD.110.124061",
    journal = "Phys. Rev. D",
    volume = "110",
    number = "12",
    pages = "124061",
    year = "2024"
}

@article{Koivisto:2026rmp,
    author = "Koivisto, Tomi S.",
    title = "{Hamiltonian surface charges in general parallel relativity}",
    eprint = "2610.16799",
    archivePrefix = "arXiv",
    primaryClass = "gr-qc",
    doi = "10.1007/s10714-026-03602-6",
    journal = "Gen. Rel. Grav.",
    volume = "58",
    number = "9",
    pages = "103",
    year = "2026"
}

@article{Koivisto:2026yxt,
    author = "Koivisto, Tomi S.",
    title = "{The canonical energy-momentum currents in cosmology}",
    eprint = "2602.06509",
    archivePrefix = "arXiv",
    primaryClass = "gr-qc",
    doi = "10.1007/s10714-026-03514-5",
    journal = "Gen. Rel. Grav.",
    volume = "58",
    number = "2",
    pages = "10",
    year = "2026"
}

@article{Bergamin:2004us,
    author = "Bergamin, L. and Grumiller, D. and Kummer, W.",
    title = "{Quantization of 2-D dilaton supergravity with matter}",
    eprint = "hep-th/0404004",
    archivePrefix = "arXiv",
    reportNumber = "TUW-04-04, LU-ITP-04-004",
    doi = "10.1088/1126-6708/2004/05/060",
    journal = "JHEP",
    volume = "05",
    pages = "060",
    year = "2004"
}

@article{Ashtekar:2005qt,
    author = "Ashtekar, Abhay and Bojowald, Martin",
    title = "{Quantum geometry and the Schwarzschild singularity}",
    eprint = "gr-qc/0509075",
    archivePrefix = "arXiv",
    reportNumber = "IGPG-05-09-01, AEI-2005-132",
    doi = "10.1088/0264-9381/23/2/008",
    journal = "Class. Quant. Grav.",
    volume = "23",
    pages = "391--411",
    year = "2006"
}

@article{Gambini:2013ooa,
    author = "Gambini, Rodolfo and Pullin, Jorge",
    title = "{Loop quantization of the Schwarzschild black hole}",
    eprint = "1302.5265",
    archivePrefix = "arXiv",
    primaryClass = "gr-qc",
    reportNumber = "LSU-REL-022113, LSU-REL-022113",
    doi = "10.1103/PhysRevLett.110.211301",
    journal = "Phys. Rev. Lett.",
    volume = "110",
    number = "21",
    pages = "211301",
    year = "2013"
}

@article{Modesto:2008im,
    author = "Modesto, Leonardo",
    title = "{Semiclassical loop quantum black hole}",
    eprint = "0811.2196",
    archivePrefix = "arXiv",
    primaryClass = "gr-qc",
    doi = "10.1007/s10773-010-0346-x",
    journal = "Int. J. Theor. Phys.",
    volume = "49",
    pages = "1649--1683",
    year = "2010"
}

@article{Gielen:2025ovv,
    author = "Gielen, Steffen and Ried, Sofie",
    title = "{Quantum Schwarzschild-(A)dS black holes: unitarity and singularity resolution}",
    eprint = "2502.10104",
    archivePrefix = "arXiv",
    primaryClass = "hep-th",
    doi = "10.1007/JHEP06(2025)074",
    journal = "JHEP",
    volume = "06",
    pages = "074",
    year = "2025"
}

@article{Davidson:2012dt,
    author = "Davidson, Aharon and Yellin, Ben",
    title = "{Schwarzschild mass uncertainty}",
    eprint = "1208.6261",
    archivePrefix = "arXiv",
    primaryClass = "gr-qc",
    doi = "10.1007/s10714-013-1662-2",
    journal = "Gen. Rel. Grav.",
    volume = "46",
    pages = "1662",
    year = "2014"
}

@article{Davidson:2014tda,
    author = "Davidson, Aharon and Yellin, Ben",
    title = "{Quantum black hole wave packet: Average area entropy and temperature dependent width}",
    eprint = "1404.5729",
    archivePrefix = "arXiv",
    primaryClass = "gr-qc",
    doi = "10.1016/j.physletb.2014.07.032",
    journal = "Phys. Lett. B",
    volume = "736",
    pages = "267--271",
    year = "2014"
}

@article{Kanatchikov:2013xmu,
    author = "Kanatchikov, I. V.",
    editor = "Diosi, Lajos and Elze, Hans-Thomas and Fronzoni, Leone and Halliwell, Jonathan and Prati, Enrico and Vitiello, Giuseppe and Yearsley, James",
    title = "{De Donder-Weyl Hamiltonian formulation and precanonical quantization of vielbein gravity}",
    eprint = "1302.2610",
    archivePrefix = "arXiv",
    primaryClass = "gr-qc",
    doi = "10.1088/1742-6596/442/1/012041",
    journal = "J. Phys. Conf. Ser.",
    volume = "442",
    pages = "012041",
    year = "2013"
}

@article{Iosifidis:2023wbx,
    author = "Iosifidis, Damianos",
    title = "{On a torsion/curvature analogue of dual connections and statistical manifolds}",
    eprint = "2303.13259",
    archivePrefix = "arXiv",
    primaryClass = "math.DG",
    doi = "10.1016/j.geomphys.2023.105064",
    journal = "J. Geom. Phys.",
    volume = "196",
    pages = "105064",
    year = "2024"
}

@article{Kanatchikov:2014nua,
    author = "Kanatchikov, I. V.",
    editor = "Kilin, Sergey Ya. and Ruffini, Remo and Vereshchagin, Gregory",
    title = "{On precanonical quantization of gravity}",
    eprint = "1407.3101",
    archivePrefix = "arXiv",
    primaryClass = "gr-qc",
    journal = "Nonlin. Phenom. Complex Syst.",
    volume = "17",
    number = "4",
    pages = "372--376",
    year = "2014"
}

@article{Kanatchikov:2016neu,
    author = "Kanatchikov, Igor V.",
    title = "{Ehrenfest Theorem in Precanonical Quantization of Fields and Gravity}",
    eprint = "1602.01083",
    archivePrefix = "arXiv",
    primaryClass = "gr-qc",
    month = "2",
    year = "2016"
}

@article{LIGOScientific:2016aoc,
    author = "Abbott, B. P. and others",
    collaboration = "LIGO Scientific, Virgo",
    title = "{Observation of Gravitational Waves from a Binary Black Hole Merger}",
    eprint = "1602.03837",
    archivePrefix = "arXiv",
    primaryClass = "gr-qc",
    reportNumber = "LIGO-P150914",
    doi = "10.1103/PhysRevLett.116.061102",
    journal = "Phys. Rev. Lett.",
    volume = "116",
    number = "6",
    pages = "061102",
    year = "2016"
}

@article{EventHorizonTelescope:2019dse,
    author = "Akiyama, Kazunori and others",
    collaboration = "Event Horizon Telescope",
    title = "{First M87 Event Horizon Telescope Results. I. The Shadow of the Supermassive Black Hole}",
    eprint = "1906.11238",
    archivePrefix = "arXiv",
    primaryClass = "astro-ph.GA",
    doi = "10.3847/2041-8213/ab0ec7",
    journal = "Astrophys. J. Lett.",
    volume = "875",
    pages = "L1",
    year = "2019"
}

@article{EventHorizonTelescope:2022wkp,
    author = "Akiyama, Kazunori and others",
    collaboration = "Event Horizon Telescope",
    title = "{First Sagittarius A* Event Horizon Telescope Results. I. The Shadow of the Supermassive Black Hole in the Center of the Milky Way}",
    eprint = "2311.08680",
    archivePrefix = "arXiv",
    primaryClass = "astro-ph.HE",
    doi = "10.3847/2041-8213/ac6674",
    journal = "Astrophys. J. Lett.",
    volume = "930",
    number = "2",
    pages = "L12",
    year = "2022"
}

@book{Polchinski:1998rq,
    author = "Polchinski, J.",
    title = "{String theory. Vol. 1: An introduction to the bosonic string}",
    doi = "10.1017/CBO9780511816079",
    isbn = "978-0-511-25227-3, 978-0-521-67227-6, 978-0-521-63303-1",
    publisher = "Cambridge University Press",
    series = "Cambridge Monographs on Mathematical Physics",
    month = "12",
    year = "2007"
}

@book{Rovelli:2004tv,
    author = "Rovelli, Carlo",
    title = "{Quantum gravity}",
    doi = "10.1017/CBO9780511755804",
    publisher = "Univ. Pr.",
    address = "Cambridge, UK",
    series = "Cambridge Monographs on Mathematical Physics",
    year = "2004"
}

@article{Ashtekar:2004eh,
    author = "Ashtekar, Abhay and Lewandowski, Jerzy",
    title = "{Background independent quantum gravity: A Status report}",
    eprint = "gr-qc/0404018",
    archivePrefix = "arXiv",
    doi = "10.1088/0264-9381/21/15/R01",
    journal = "Class. Quant. Grav.",
    volume = "21",
    pages = "R53",
    year = "2004"
}

@article{Ambjorn:2012jv,
    author = "Ambjorn, J. and Goerlich, A. and Jurkiewicz, J. and Loll, R.",
    title = "{Nonperturbative Quantum Gravity}",
    eprint = "1203.3591",
    archivePrefix = "arXiv",
    primaryClass = "hep-th",
    doi = "10.1016/j.physrep.2012.03.007",
    journal = "Phys. Rept.",
    volume = "519",
    pages = "127--210",
    year = "2012"
}

@article{Reuter:2012id,
    author = "Reuter, Martin and Saueressig, Frank",
    title = "{Quantum Einstein Gravity}",
    eprint = "1202.2274",
    archivePrefix = "arXiv",
    primaryClass = "hep-th",
    doi = "10.1088/1367-2630/14/5/055022",
    journal = "New J. Phys.",
    volume = "14",
    pages = "055022",
    year = "2012"
}

@article{Koch:2016uso,
    author = "Koch, Benjamin and Reyes, Ignacio A. and Rinc{\'o}n, {\'A}ngel",
    title = "{A scale dependent black hole in three-dimensional space{\textendash}time}",
    eprint = "1606.04123",
    archivePrefix = "arXiv",
    primaryClass = "hep-th",
    doi = "10.1088/0264-9381/33/22/225010",
    journal = "Class. Quant. Grav.",
    volume = "33",
    number = "22",
    pages = "225010",
    year = "2016"
}

@article{Rincon:2017goj,
    author = "Rinc{\'o}n, {\'A}ngel and Contreras, Ernesto and Bargue{\~n}o, Pedro and Koch, Benjamin and Panotopoulos, Grigorios and Hern{\'a}ndez-Arboleda, Alejandro",
    title = "{Scale dependent three-dimensional charged black holes in linear and non-linear electrodynamics}",
    eprint = "1704.04845",
    archivePrefix = "arXiv",
    primaryClass = "hep-th",
    doi = "10.1140/epjc/s10052-017-5045-9",
    journal = "Eur. Phys. J. C",
    volume = "77",
    number = "7",
    pages = "494",
    year = "2017"
}

@article{Szabo:2001kg,
    author = "Szabo, Richard J.",
    title = "{Quantum field theory on noncommutative spaces}",
    eprint = "hep-th/0109162",
    archivePrefix = "arXiv",
    reportNumber = "HWM-01-35, EMPG-01-14",
    doi = "10.1016/S0370-1573(03)00059-0",
    journal = "Phys. Rept.",
    volume = "378",
    pages = "207--299",
    year = "2003"
}

@book{Connes:1994yd,
    author = "Connes, Alain",
    title = "{Noncommutative geometry}",
    isbn = "978-0-12-185860-5",
    year = "1994"
}

@article{Shenderov:2024rup,
    author = "Shenderov, Victoria and Kanex, Mark and Beitel, Thomas and Tobar, Germain and Manikandan, Sreenath K. and Pikovski, Igor",
    title = "{Stimulated absorption of single gravitons: First light on quantum gravity}",
    eprint = "2407.11929",
    archivePrefix = "arXiv",
    primaryClass = "gr-qc",
    doi = "10.1016/j.aop.2026.170448",
    journal = "Annals Phys.",
    volume = "489",
    pages = "170448",
    year = "2026"
}

@article{Hehl:1994ue,
    author = "Hehl, Friedrich W. and McCrea, J. Dermott and Mielke, Eckehard W. and Ne'eman, Yuval",
    title = "{Metric affine gauge theory of gravity: Field equations, Noether identities, world spinors, and breaking of dilation invariance}",
    eprint = "gr-qc/9402012",
    archivePrefix = "arXiv",
    reportNumber = "TAUP-N192-94, TAUP-192-94",
    doi = "10.1016/0370-1573(94)00111-F",
    journal = "Phys. Rept.",
    volume = "258",
    pages = "1--171",
    year = "1995"
}

@article{Freidel:2026hed,
    author = "Freidel, Laurent and Oberfrank, Robin",
    title = "{Geometric noise spectrum in interferometers}",
    eprint = "2601.17849",
    archivePrefix = "arXiv",
    primaryClass = "hep-th",
    month = "1",
    year = "2026"
}

@article{Calmet:2017qqa,
    author = "Calmet, Xavier and El-Menoufi, Basem Kamal",
    title = "{Quantum Corrections to Schwarzschild Black Hole}",
    eprint = "1704.00261",
    archivePrefix = "arXiv",
    primaryClass = "hep-th",
    doi = "10.1140/epjc/s10052-017-4802-0",
    journal = "Eur. Phys. J. C",
    volume = "77",
    number = "4",
    pages = "243",
    year = "2017"
}

@article{Donoghue:1994dn,
    author = "Donoghue, John F.",
    title = "{General relativity as an effective field theory: The leading quantum corrections}",
    eprint = "gr-qc/9405057",
    archivePrefix = "arXiv",
    reportNumber = "UMHEP-408",
    doi = "10.1103/PhysRevD.50.3874",
    journal = "Phys. Rev. D",
    volume = "50",
    pages = "3874--3888",
    year = "1994"
}

@article{Thiemann:1992jj,
    author = "Thiemann, T. and Kastrup, H. A.",
    title = "{Canonical quantization of spherically symmetric gravity in Ashtekar's selfdual representation}",
    eprint = "gr-qc/9310012",
    archivePrefix = "arXiv",
    reportNumber = "PITHA-92-23",
    doi = "10.1016/0550-3213(93)90623-W",
    journal = "Nucl. Phys. B",
    volume = "399",
    pages = "211--258",
    year = "1993"
}

@article{Heisenberg:2026ess,
    author = "Heisenberg, Lavinia",
    title = "{Autoparallels and the Inverse Problem of the Calculus of Variations}",
    eprint = "2603.11581",
    archivePrefix = "arXiv",
    primaryClass = "math-ph",
    doi = "10.1002/prop.70115",
    journal = "Fortsch. Phys.",
    volume = "74",
    number = "5",
    pages = "e70115",
    year = "2026"
}

@article{Csillag:2026kdc,
    author = "Csillag, Lehel and Voicu, Nicoleta and Elgendi, Salah and Pfeifer, Christian",
    title = "{On the Finsler variational nature of autoparallels in metric-affine geometry}",
    eprint = "2603.18416",
    archivePrefix = "arXiv",
    primaryClass = "math-ph",
    month = "3",
    year = "2026"
}

@article{Dirac:1950pj,
    author = "Dirac, Paul A. M.",
    title = "{Generalized Hamiltonian dynamics}",
    doi = "10.4153/CJM-1950-012-1",
    journal = "Can. J. Math.",
    volume = "2",
    pages = "129--148",
    year = "1950"
}

@book{dirac2013lectures,
	author = {Dirac, P.A.M.},
	isbn = {9780486320281},
	publisher = {Dover Publications},
	series = {Dover Books on Physics},
	title = {Lectures on Quantum Mechanics},
	url = {https://books.google.at/books?id=Z3XCAgAAQBAJ},
	year = {2013}}

@inproceedings{Date:2010xr,
    author = "Date, Ghanashyam",
    title = "{Lectures on Constrained Systems}",
    booktitle = "{Refresher Course for College Teachers}",
    eprint = "1010.2062",
    archivePrefix = "arXiv",
    primaryClass = "gr-qc",
    month = "10",
    year = "2010"
}

@article{Armaza:2016vfz,
    author = "Armaza, Crist{\'o}bal and Hojman, Sergio A. and Koch, Benjamin and Zalaquett, Nicol{\'a}s",
    title = "{On the possibility of non-geodesic motion of massless spinning tops}",
    eprint = "1601.05809",
    archivePrefix = "arXiv",
    primaryClass = "gr-qc",
    doi = "10.1088/0264-9381/33/14/145011",
    journal = "Class. Quant. Grav.",
    volume = "33",
    number = "14",
    pages = "145011",
    year = "2016"
}

@article{Gambini:1998it,
    author = "Gambini, Rodolfo and Pullin, Jorge",
    title = "{Nonstandard optics from quantum space-time}",
    eprint = "gr-qc/9809038",
    archivePrefix = "arXiv",
    reportNumber = "CGPG-98-9-1",
    doi = "10.1103/PhysRevD.59.124021",
    journal = "Phys. Rev. D",
    volume = "59",
    pages = "124021",
    year = "1999"
}

@article{Asenjo:2016hnc,
    author = "Asenjo, Felipe A. and Hojman, Sergio A.",
    title = "{Do electromagnetic waves always propagate along null geodesics?}",
    eprint = "1608.06572",
    archivePrefix = "arXiv",
    primaryClass = "gr-qc",
    doi = "10.1088/1361-6382/aa8b48",
    journal = "Class. Quant. Grav.",
    volume = "34",
    number = "20",
    pages = "205011",
    year = "2017"
}

@article{Parikh:2020kfh,
    author = "Parikh, Maulik and Wilczek, Frank and Zahariade, George",
    title = "{Quantum Mechanics of Gravitational Waves}",
    eprint = "2010.08205",
    archivePrefix = "arXiv",
    primaryClass = "hep-th",
    doi = "10.1103/PhysRevLett.127.081602",
    journal = "Phys. Rev. Lett.",
    volume = "127",
    number = "8",
    pages = "081602",
    year = "2021"
}

@article{Dalvit:1999wd,
    author = "Dalvit, Diego A. R. and Mazzitelli, Francisco D.",
    title = "{Quantum corrected geodesics}",
    eprint = "gr-qc/9904051",
    archivePrefix = "arXiv",
    doi = "10.1103/PhysRevD.60.084018",
    journal = "Phys. Rev. D",
    volume = "60",
    pages = "084018",
    year = "1999"
}

@article{Piazza:2022amf,
    author = "Piazza, Federico and Tolley, Andrew J.",
    title = "{Subadditive average distances and quantum promptness}",
    eprint = "2212.06156",
    archivePrefix = "arXiv",
    primaryClass = "hep-th",
    reportNumber = "Imperial/TP/2022/AJT/02",
    doi = "10.1088/1361-6382/ace583",
    journal = "Class. Quant. Grav.",
    volume = "40",
    number = "16",
    pages = "165013",
    year = "2023"
}

@article{Kanatchikov:2025tlm,
    author = "Kanatchikov, Igor V. and Kholodnyi, Valery A.",
    title = "{Observable signatures of precanonical quantum gravity}",
    doi = "10.1209/0295-5075/addcdb",
    journal = "EPL",
    volume = "150",
    number = "5",
    pages = "59002",
    year = "2025"
}

@article{Pipa:2018bui,
    author = "Pipa, Francisco and Paunkovi{\'c}, Nikola and Vojinovi{\'c}, Marko",
    title = "{Entanglement-induced deviation from the geodesic motion in quantum gravity}",
    eprint = "1801.03207",
    archivePrefix = "arXiv",
    primaryClass = "gr-qc",
    doi = "10.1088/1475-7516/2019/09/057",
    journal = "JCAP",
    volume = "09",
    pages = "057",
    year = "2019"
}

@article{Saravani:2015rva,
    author = "Saravani, Mehdi and Aslanbeigi, Siavash",
    title = "{Dark Matter From Spacetime Nonlocality}",
    eprint = "1502.01655",
    archivePrefix = "arXiv",
    primaryClass = "hep-th",
    doi = "10.1103/PhysRevD.92.103504",
    journal = "Phys. Rev. D",
    volume = "92",
    number = "10",
    pages = "103504",
    year = "2015"
}

@article{Amelino-Camelia:2011lvm,
    author = "Amelino-Camelia, Giovanni and Freidel, Laurent and Kowalski-Glikman, Jerzy and Smolin, Lee",
    title = "{The principle of relative locality}",
    eprint = "1101.0931",
    archivePrefix = "arXiv",
    primaryClass = "hep-th",
    doi = "10.1103/PhysRevD.84.084010",
    journal = "Phys. Rev. D",
    volume = "84",
    pages = "084010",
    year = "2011"
}

@article{Belenchia:2014fda,
    author = "Belenchia, Alessio and Benincasa, Dionigi M. T. and Liberati, Stefano",
    title = "{Nonlocal Scalar Quantum Field Theory from Causal Sets}",
    eprint = "1411.6513",
    archivePrefix = "arXiv",
    primaryClass = "gr-qc",
    doi = "10.1007/JHEP03(2015)036",
    journal = "JHEP",
    volume = "03",
    pages = "036",
    year = "2015"
}

@article{Einstein:1916vd,
    author = "Einstein, Albert",
    editor = "Hsu, Jong-Ping and Fine, D.",
    title = "{The foundation of the general theory of relativity.}",
    doi = "10.1002/andp.19163540702",
    journal = "Annalen Phys.",
    volume = "49",
    number = "7",
    pages = "769--822",
    year = "1916"
}

@article{Einstein:632328,
      author        = "Einstein, A",
      title         = "{Näherungsweise Integration der Feldgleichungen der
                       Gravitation, 22 Jun 1916}",
      year          = "1916",
      url           = "https://cds.cern.ch/record/632328",
}

@article{Einstein:1915ca,
    author  = {Einstein, Albert},
    title   = {Die Feldgleichungen der Gravitation},
    journal = {Sitzungsberichte der K{\"o}niglich Preussischen Akademie der Wissenschaften},
    pages   = {844--847},
    year    = {1915}
}

@article{Schwarzschild:1916uq,
    author  = {Schwarzschild, Karl},
    title   = {{\"U}ber das Gravitationsfeld eines Massenpunktes nach der Einsteinschen Theorie},
    journal = {Sitzungsberichte der K{\"o}niglich Preussischen Akademie der Wissenschaften},
    pages   = {189--196},
    year    = {1916}
}

@article{Heisenberg:1925,
    author  = {Heisenberg, Werner},
    title   = {{\"U}ber quantentheoretische Umdeutung kinematischer und mechanischer Beziehungen},
    journal = {Zeitschrift f{\"u}r Physik},
    volume  = {33},
    pages   = {879--893},
    year    = {1925},
    doi     = {10.1007/BF01328377}
}

@article{Schrodinger:1926,
    author  = {Schr{\"o}dinger, Erwin},
    title   = {Quantisierung als Eigenwertproblem},
    journal = {Annalen der Physik},
    volume  = {79},
    pages   = {361--376},
    year    = {1926},
    doi     = {10.1002/andp.19263840404}
}

@article{vonNeumann:1927,
    author  = {von Neumann, John},
    title   = {Mathematische Begr{\"u}ndung der Quantenmechanik},
    journal = {Nachrichten von der Gesellschaft der Wissenschaften zu G{\"o}ttingen,
               Mathematisch-Physikalische Klasse},
    volume  = {1927},
    pages   = {1--57},
    year    = {1927}
}

@article{Dirac:1927,
    author  = {Dirac, Paul A. M.},
    title   = {The Quantum Theory of the Emission and Absorption of Radiation},
    journal = {Proceedings of the Royal Society of London A},
    volume  = {114},
    number  = {767},
    pages   = {243--265},
    year    = {1927},
    doi     = {10.1098/rspa.1927.0039}
}

@article{Killing:1892,
    author  = {Killing, Wilhelm},
    title   = {Ueber die Grundlagen der Geometrie},
    journal = {Journal f{\"u}r die reine und angewandte Mathematik},
    volume  = {109},
    pages   = {121--186},
    year    = {1892},
    doi     = {10.1515/crll.1892.109.121}
}

@article{Slebodzinski:1931,
    author  = {{\'S}lebodzi{\'n}ski, W{\l}adys{\l}aw},
    title   = {Sur les {\'e}quations de Hamilton},
    journal = {Bulletin de l'Acad{\'e}mie Royale de Belgique},
    series  = {5},
    volume  = {17},
    pages   = {864--870},
    year    = {1931}
}

@article{Wigner:1939,
    author  = {Wigner, Eugene P.},
    title   = {On Unitary Representations of the Inhomogeneous Lorentz Group},
    journal = {Annals of Mathematics},
    volume  = {40},
    number  = {1},
    pages   = {149--204},
    year    = {1939},
    doi     = {10.2307/1968551}
}

@article{Stone:1932,
    author  = {Stone, Marshall H.},
    title   = {On One-Parameter Unitary Groups in Hilbert Space},
    journal = {Annals of Mathematics},
    volume  = {33},
    number  = {3},
    pages   = {643--648},
    year    = {1932},
    doi     = {10.2307/1968538}
}

@article{Will:2014,
    author  = {Will, Clifford M.},
    title   = {The Confrontation between General Relativity and Experiment},
    journal = {Living Reviews in Relativity},
    volume  = {17},
    number  = {4},
    year    = {2014},
    doi     = {10.12942/lrr-2014-4},
    eprint  = {1403.7377},
    archivePrefix = {arXiv},
    primaryClass  = {gr-qc}
}

@article{Noether:1918,
    author  = {Noether, Emmy},
    title   = {Invariante Variationsprobleme},
    journal = {Nachrichten von der Gesellschaft der Wissenschaften zu G{\"o}ttingen,
               Mathematisch-Physikalische Klasse},
    volume  = {1918},
    pages   = {235--257},
    year    = {1918}
}

@article{Matschull:1996up,
    author = "Matschull, Hans-Juergen",
    title = "{Dirac's Canonical Quantization Programme}",
    eprint = "quant-ph/9606031",
    archivePrefix = "arXiv",
    month = "6",
    year = "1996"
}

@article{Wipf:1993xg,
    author = "Wipf, Andreas W.",
    editor = "Ehlers, J. and Friedrich, H.",
    title = "{Hamilton's formalism for systems with constraints}",
    eprint = "hep-th/9312078",
    archivePrefix = "arXiv",
    reportNumber = "ETH-TH-93-48",
    doi = "10.1007/3-540-58339-4_14",
    journal = "Lect. Notes Phys.",
    volume = "434",
    pages = "22--58",
    year = "1994"
}

@article{Faddeev:1988qp,
    author = "Faddeev, L. D. and Jackiw, R.",
    title = "{Hamiltonian Reduction of Unconstrained and Constrained Systems}",
    reportNumber = "MIT-CTP-1567",
    doi = "10.1103/PhysRevLett.60.1692",
    journal = "Phys. Rev. Lett.",
    volume = "60",
    pages = "1692--1694",
    year = "1988"
}

@inproceedings{Gitman:2004qs,
    author = "Gitman, D. M. and Tyutin, I. V.",
    title = "{Symmetries in constrained systems}",
    booktitle = "{2nd Conference on Lie and Jordan Algebras, their Representations and Applications}",
    eprint = "hep-th/0409087",
    archivePrefix = "arXiv",
    reportNumber = "IFUSP-1257-2004",
    month = "9",
    year = "2004"
}

@article{Castellani:1981us,
    author = "Castellani, Leonardo",
    title = "{Symmetries in Constrained Hamiltonian Systems}",
    reportNumber = "ITP-SB-81-5",
    doi = "10.1016/0003-4916(82)90031-8",
    journal = "Annals Phys.",
    volume = "143",
    pages = "357",
    year = "1982"
}

@article{RicciLeviCivita:1900,
    author  = {Ricci-Curbastro, Gregorio and Levi-Civita, Tullio},
    title   = {M{\'e}thodes de calcul diff{\'e}rentiel absolu
               et leurs applications},
    journal = {Mathematische Annalen},
    volume  = {54},
    pages   = {125--201},
    year    = {1901},
    doi     = {10.1007/BF01454201}
}

@article{Jebsen:1921,
    author  = {Jebsen, J. T.},
    title   = {{\"U}ber die allgemeinen kugelsymmetrischen L{\"o}sungen
               der Einsteinschen Gravitationsgleichungen im Vakuum},
    journal = {Arkiv f{\"o}r Matematik, Astronomi och Fysik},
    volume  = {15},
    number  = {18},
    pages   = {1--9},
    year    = {1921}
}

@article{Eiesland:1925,
    author  = {Eiesland, John A.},
    title   = {The Group of Motions of an Einstein Space},
    journal = {Transactions of the American Mathematical Society},
    volume  = {27},
    number  = {2},
    pages   = {213--245},
    year    = {1925},
    doi     = {10.1090/S0002-9947-1925-1501308-7}
}

@article{Poisson:2011nh,
    author        = {Poisson, Eric and Pound, Adam and Vega, Ian},
    title         = {The Motion of Point Particles in Curved Spacetime},
    journal       = {Living Reviews in Relativity},
    volume        = {14},
    pages         = {7},
    year          = {2011},
    doi           = {10.12942/lrr-2011-7},
    eprint        = {1102.0529},
    archivePrefix = {arXiv},
    primaryClass  = {gr-qc}
}

@article{Droste:1917,
    author  = {Droste, Johannes},
    title   = {The Field of a Single Centre in Einstein's Theory of
               Gravitation, and the Motion of a Particle in That Field},
    journal = {Proceedings of the Royal Netherlands Academy of Arts
               and Sciences},
    volume  = {19},
    pages   = {197--215},
    year    = {1917}
}

@article{Hagihara:1931,
    author  = {Hagihara, Yusuke},
    title   = {Theory of the Relativistic Trajectories in a
               Gravitational Field of Schwarzschild},
    journal = {Japanese Journal of Astronomy and Geophysics},
    volume  = {8},
    pages   = {67--176},
    year    = {1931}
}

@article{Koch:2025qzv,
    author = "Koch, Benjamin and Riahinia, Ali and Rincon, Angel",
    title = "{Geodesics in quantum gravity}",
    eprint = "2510.00117",
    archivePrefix = "arXiv",
    primaryClass = "gr-qc",
    doi = "10.1103/w1sd-v69d",
    journal = "Phys. Rev. D",
    volume = "112",
    number = "8",
    pages = "084056",
    year = "2025"
}

@article{Koch:2025yuz,
    author = "Koch, Benjamin and Riahinia, Ali",
    title = "{Quantum Uncertainties of Static Spherically Symmetric Spacetimes}",
    eprint = "2506.22545",
    archivePrefix = "arXiv",
    primaryClass = "gr-qc",
    doi = "10.3390/universe12030059",
    journal = "Universe",
    volume = "12",
    number = "3",
    pages = "59",
    year = "2026"
}

@article{Giacomini2019QRF,
  author  = {Giacomini, Flaminia and Castro-Ruiz, Esteban and Brukner, {\v C}aslav},
  title   = {Quantum mechanics and the covariance of physical laws in quantum reference frames},
  journal = {Nature Communications},
  volume  = {10},
  pages   = {494},
  year    = {2019},
  doi     = {10.1038/s41467-018-08155-0}
}

@article{Vanrietvelde2020Perspective,
  author  = {Vanrietvelde, Augustin and H{\"o}hn, Philipp A. and Giacomini, Flaminia and Castro-Ruiz, Esteban},
  title   = {A change of perspective: switching quantum reference frames via a perspective-neutral framework},
  journal = {Quantum},
  volume  = {4},
  pages   = {225},
  year    = {2020},
  doi     = {10.22331/q-2020-01-27-225}
}

@article{deLaHamette2023IndefiniteMetric,
  author  = {de la Hamette, Anne-Catherine and Kabel, Viktoria and Castro-Ruiz, Esteban and Brukner, {\v C}aslav},
  title   = {Quantum reference frames for an indefinite metric},
  journal = {Communications Physics},
  volume  = {6},
  pages   = {231},
  year    = {2023},
  doi     = {10.1038/s42005-023-01344-4}
}

@article{Kabel2023Boundary,
  author  = {Kabel, Viktoria and Brukner, {\v C}aslav and Wieland, Wolfgang},
  title   = {Quantum reference frames at the boundary of spacetime},
  journal = {Physical Review D},
  volume  = {108},
  pages   = {106022},
  year    = {2023},
  doi     = {10.1103/PhysRevD.108.106022}
}

@misc{HausmannRenner2026ThirdParticle,
  author        = {Hausmann, Ladina and Renner, Renato},
  title         = {The Paradox of the Third Particle is classical},
  year          = {2026},
  eprint        = {2607.15351},
  archivePrefix = {arXiv},
  primaryClass  = {quant-ph}
}

@article{DeWitt1967,
  author  = {DeWitt, Bryce S.},
  title   = {Quantum Theory of Gravity. I. The Canonical Theory},
  journal = {Physical Review},
  volume  = {160},
  pages   = {1113--1148},
  year    = {1967},
  doi     = {10.1103/PhysRev.160.1113}
}

@inproceedings{Kuchar1993Canonical,
  author    = {Kucha{\v r}, Karel V.},
  title     = {Canonical Quantum Gravity},
  booktitle = {General Relativity and Gravitation 1992:
               Proceedings of the Thirteenth International Conference
               on General Relativity and Gravitation},
  editor    = {Gleiser, R. J. and Kozameh, C. N. and Moreschi, O. M.},
  pages     = {119--150},
  publisher = {Institute of Physics Publishing},
  address   = {Bristol},
  year      = {1993},
  eprint    = {gr-qc/9304012},
  archivePrefix = {arXiv}
}

@book{Thiemann2007,
  author    = {Thiemann, Thomas},
  title     = {Modern Canonical Quantum General Relativity},
  publisher = {Cambridge University Press},
  year      = {2007},
  doi       = {10.1017/CBO9780511755682}
}

@article{BeigChrusciel1997KID,
  author  = {Beig, Robert and Chru{\'s}ciel, Piotr T.},
  title   = {Killing Initial Data},
  journal = {Classical and Quantum Gravity},
  volume  = {14},
  number  = {1A},
  pages   = {A83--A92},
  year    = {1997},
  doi     = {10.1088/0264-9381/14/1A/007},
  eprint  = {gr-qc/9604040},
  archivePrefix = {arXiv}
}
\end{document}